\documentclass[default]{sn-jnl}

\usepackage[acronyms,nonumberlist,automake=true,nogroupskip,nopostdot]{glossaries} 
\usepackage{gensymb}
\usepackage{bbm}

\usepackage{graphicx}%
\usepackage{multirow}%
\usepackage{amsmath,amssymb,amsfonts}%
\usepackage{amsthm}%
\usepackage{mathrsfs}%
\usepackage[title]{appendix}%
\usepackage{xcolor}%
\usepackage{textcomp}%
\usepackage{manyfoot}%
\usepackage{booktabs}%
\usepackage{algorithm}%
\usepackage{algorithmicx}%
\usepackage{algpseudocode}%
\usepackage{listings}%
\usepackage{verbatim}

\theoremstyle{thmstyleone}%
\theoremstyle{thmstyletwo}%

\theoremstyle{thmstylethree}%

\makeglossaries 

\begin{document}
\title[Article Title]{\textbf{Extremes of summer Arctic sea ice reduction investigated with a rare event algorithm}}

\author*[1]{\textbf{\fnm{Jerome} \sur{Sauer}}}\email{jerome.sauer@uclouvain.be}

\author[2]{\textbf{\fnm{Jonathan} \sur{Demaeyer}}}

\author[3]{\textbf{\fnm{Giuseppe} \sur{Zappa}}}

\author[1]{\textbf{\fnm{François} \sur{Massonnet}}}

\author[1,2]{\textbf{\fnm{Francesco}} \textbf{\sur{Ragone}}}

\affil*[1]{\orgdiv{Earth and Climate Research Center, Earth and Life Institute}, \orgname{Université catholique de Louvain}, \orgaddress{\city{Louvain-la-Neuve}, \country{Belgium}}}

\affil[2]{\orgdiv{Royal Meteorological Institute of Belgium}, \city{Brussels}, \country{Belgium}}

\affil[3]{\orgdiv{National Research Council of Italy}, \orgname{Institute of Atmospheric Sciences and Climate}, \orgaddress{\city{Bologna}, \country{Italy}}}


\abstract{Various studies identified possible drivers of extremes of Arctic sea ice reduction, such as observed in the summers of 2007 and 2012, including preconditioning, local feedback mechanisms, oceanic heat transport and the synoptic- and large-scale atmospheric circulations. However, a robust quantitative statistical analysis of extremes of sea ice reduction is hindered by the small number of events that can be sampled in observations and numerical simulations with computationally expensive climate models. Recent studies tackled the problem of sampling climate extremes by using rare event algorithms, i.e., computational techniques developed in statistical physics to reduce the computational cost required to sample rare events in numerical simulations. Here we apply a rare event algorithm to ensemble simulations with the intermediate complexity coupled climate model PlaSim-LSG to investigate extreme negative summer pan-Arctic sea ice area anomalies under pre-industrial greenhouse gas conditions. Owing to the algorithm, we estimate return times of extremes orders of magnitude larger than feasible with direct sampling, and we compute statistically significant composite maps of dynamical quantities conditional on the occurrence of these extremes. We find that extremely low sea ice summers in PlaSim-LSG are associated with preconditioning through the winter sea ice-ocean state, with enhanced downward longwave radiation due to an anomalously moist and warm spring Arctic atmosphere and with enhanced downward sensible heat fluxes during the spring-summer transition. As a consequence of these three processes, the sea ice-albedo feedback becomes active in spring and leads to an amplification of pre-existing sea ice area anomalies during summer.}

\keywords{Rare event algorithms, Climate extremes, Arctic sea ice variability, Climate model, EMIC}

\maketitle
\newpage   

\section{Introduction}\label{sec1}

The Arctic sea ice cover has been shrinking since at least the late 1970s \citep{notz2012,stroeve2018}, in large part due to anthropogenic emissions of greenhouse gases \citep{gregory2002,notz2012,stroeve2018,ding2017}. On top of the downward trend, internal climate variability contributes to the year-to-year variations and associated extreme events of the annual sea ice minimum in September \citep{francis2020,ono2019mechanisms}.

Extreme Arctic sea ice reduction, such as observed in the summers of 2007 and 2012, may have impacts outside the Arctic Ocean. Studies have suggested a possible influence of sea ice loss on the integrity of the permafrost \citep{lawrence2008} and on weather patterns and climate in the mid to high latitudes \citep{petoukhov2010,francis2009winter, francis2012,screen2018,petrie2015atmospheric,chripko2021,delhaye2022}. Apart from its effect on the climate system, Arctic sea ice decline leads to increased marine accessibility. This has implications for the opening of trans-Arctic shipping routes, offshore industries, polar ecotourism and the daily life of local communities \citep{eicken2013,smith2013,lloyd2012}.  
 
Possible drivers of extreme summer Arctic sea ice reduction have been suggested in the literature. Both the extreme sea ice lows in 2007 and 2012 were attributed to climate change via preconditioning through the ongoing winter sea ice thinning and to the occurrence of particular weather and climate events \citep{lindsay2009,kauker2009adjoint,zhang2008,zhang2013,parkinson2013}. In 2007, sea ice reduction was favoured by enhanced inflow of warm Pacific water through Bering Strait \citep{woodgate2010} and by anomalously persistent southerly winds in the Pacific sector associated with the \newacronym{ad}{AD}{Arctic Dipole} \acrfull{ad} pattern \citep{wang2009dipole,lindsay2009,overland2012,kauker2009adjoint}. In 2012, a summer storm contributed to enhanced sea ice reduction by leading to increased bottom melt via anomalously strong vertical mixing in the oceanic boundary layer \citep{guemas2013,zhang2013}. Further possible triggers of anomalously low summer Arctic sea ice area include enhanced North Atlantic oceanic heat transport \citep{aarthun2012}, the positive and negative phases of the winter and summer \newacronym{ao}{AO}{Arctic Oscillation} \acrfull{ao} \citep[e.g.][]{rigor2002response,ogi2016}, reduced cloudiness during summer \citep{schweiger2008}, and increased surface downward longwave radiation related to enhanced poleward atmospheric moisture transport during spring \citep{kapsch2013,kapsch2019}.

Even though different physical drivers have been suggested to contribute to individual extremes of Arctic sea ice reduction, a quantitative statistical analysis of their physical drivers is hindered by the poor sampling of extreme events in observations and in numerical simulations. The record of satellite-based sea ice observations includes only a few annual sea ice minima with orders of magnitude comparable to the ones in September 2007 and 2012 \citep{fetterer2017}. Moreover, the large computational cost of state-of-the-art general circulation models makes it unrealistic to run them longer than a few thousands of years and to quantitatively study sea ice extremes with return times of longer than order 10\textsuperscript{2} years. Going beyond individual case studies, a generalized analysis of the relative importance of different atmospheric and oceanic precursors of extremes of sea ice reduction remains therefore a challenge. A better understanding of the precursors of extremes of summer Arctic sea ice reduction and a more precise estimate of their probabilities are in turn crucial to improve seasonal predictions of these events and to assess their risk of occurrence under different climate change scenarios.

In this work, we address the problem of computational cost limitations by using a rare event algorithm. Rare event algorithms are computational techniques developed in statistical physics to improve the sampling efficiency of rare events in numerical simulations \citep[e.g.][]{ragone2018,ragone2019,ragone2021}. Compared to conventional numerical simulations with the same computational cost, rare event algorithms enable to increase the number of simulated extreme events by several orders of magnitude. In this way, these techniques allow to reduce the uncertainty of return time estimates and of conditional statistics on extreme events (e.g. composites) compared to conventional simulation strategies, and to generate ultra-rare events that are very unlikely to be observed using direct sampling. Rare event algorithms have been introduced in the 1950s \citep{kahn1951} and have been used since for a wide range of applications (for an overview and the mathematical analysis see e.g. \cite{delmoral2004,giardina2011simulating,grafke2019}). Recently, some of these techniques have been applied in climate science and in fluid dynamics to study heat waves \citep{ragone2018,ragone2019,ragone2021}, midlatitude precipitation \citep{wouters2023rare}, tropical storms \citep{plotkin2019,webber2019}, weakening and collapse of the  \newacronym{amoc}{AMOC}{Atlantic meridional overturning circulation} \acrfull{amoc} \citep{cini2023}, 
and turbulence \citep{bouchet2018fluctuations,grafke2015instanton,lestang2020numerical}.

Apart from rare event algorithms, other techniques exist that allow to study extreme events with return times beyond the range of available data. For example, \newacronym{evt}{EVT}{extreme value theory} \acrfull{evt} can be used to extrapolate return times for anomaly values unavailable from the data set \citep{coles2001,parey2010}. \acrshort{evt} is used in climate science in particular for attribution studies, as it is an efficient way to estimate probabilities of events outside the range of observational data. However, one of the drawbacks is that it does not provide the dynamics that leads to the extrapolated extreme events. Moreover, a recently published approach to focus the computational power in climate model simulations on trajectories that lead to extreme events is ensemble boosting \citep{fischer2023,gessner2023}. In this approach, ensemble simulations are re-initialized days to weeks before the peak of an extreme event allowing to generate physically plausible ultra-rare events with unprecedented amplitude. Ensemble boosting thus allows to study the dynamics of unprecedented extreme events, but it does not provide the probabilities of these events and can therefore be used only in a story-line sense. Rare event algorithms instead deliver both the dynamical trajectories that lead to the extreme events of interest and enable to estimate their probabilities of occurrence. This allows for example to explore covariates of an observable of interest without the need of assumptions on the underlying distributions.

Different types of rare event algorithms are suited to study different problems, as the underlying design mechanisms are tailored to target extremes with specific properties. Here we use a genealogical selection algorithm \citep{ragone2018,ragone2019,ragone2021} adapted from \cite{del2005genealogical,giardina2011simulating}, that is appropriate to study persistent, long lasting events. We apply this rare event algorithm to the intermediate complexity coupled climate model PlaSim-LSG \citep{fraedrich2005,drijfhout1996,maier1993} to investigate extreme negative summer pan-Arctic sea ice area anomalies under fixed pre-industrial greenhouse gas conditions. Owing to the algorithm, we compute return times two to three orders of magnitude larger than feasible with direct sampling and we obtain statistically significant composite maps of dynamical quantities conditional on the occurrence of extremely low sea ice summers with return times of more than 200 years. Finally, we elaborate possible physical drivers of these events.  
The paper is structured as follows. In section 2, we present the set-up of the model and the control run and we describe the methodology of the rare event algorithm and the design of the rare event algorithm experiments. In section 3, we show that the rare event algorithm performs importance sampling of summer seasons with extremely low pan-Arctic sea ice area and we discuss possible physical drivers of these events. In this context, we describe the average states of the atmosphere and of the sea ice during extremely low sea ice summers, we discuss the relative contribution of winter preconditoning vs. intra-seasonal sea ice reduction to extreme negative summer pan-Arctic sea ice area anomalies and we perform a surface energy budget analysis in order to trace back extremely low sea ice conditions to anomalous atmospheric conditions. In section 4, we present our conclusions and we discuss possible future lines of research.

\section{Materials and methods}\label{sec2}
\subsection{The Planet Simulator with a Large-Scale Geostrophic ocean circulation model}\label{subsec1}

All the following simulations are conducted with the intermediate complexity climate model \newacronym{plasim}{PlaSim}{Planet Simulator} \acrfull{plasim} version 17 \citep{fraedrich2005}. To account for the importance of oceanic processes for Arctic sea ice variability, we use a coupled version of \acrshort{plasim} that includes a dynamic \newacronym{lsg}{LSG}{Large-Scale Geostrophic} \acrfull{lsg} ocean \citep{maier1993,drijfhout1996} in addition to a mixed-layer ocean model and to a thermodynamic sea ice model (\acrshort{plasim} coupled to \acrshort{lsg} is referred to as \newacronym{plasim-lsg}{PlaSim-LSG}{Planet Simulator Large-Scale Geostrophic} "\acrshort{plasim-lsg}" in this work). \acrshort{plasim-lsg} is computationally less demanding than Earth system models used in the assessments of the \newacronym{ipcc}{IPCC}{Intergovernmental Panel on Climate Change} \acrfull{ipcc}. It allows to conduct large ensemble simulations at reasonable computational cost, which is suitable for the aim of this work to investigate the applicability of a rare event algorithm to study extremes of Arctic sea ice reduction with a numerical climate model. Despite its intermediate complexity, \acrshort{plasim-lsg} produces a reasonably realistic present-day climate, especially in the northern hemisphere \citep{angeloni2022}. Further applications for which \acrshort{plasim-lsg} has been used include aquaplanet and paleoclimate studies \citep{hertwig,andres2019towards} and a study on the linkage between high-latitude precipitation and low-frequency variability of the \acrshort{amoc} \citep{mehling2023}.

The \acrshort{plasim-lsg} atmosphere includes a wet primitive equation dynamical core governing the conservation of momentum, mass, energy, the specific humidity and the equation of state with hydrostatic approximation \citep{fraedrich2005,lunkeit2012}. The equations are solved on a terrain following $\sigma$-coordinate system (pressure divided by surface pressure) with a spectral transform method \citep{orszag,eliassen}. Parameterizations of unresolved subgrid scale processes include moist and dry convection \citep{kuo1965,kuo1974}, clouds \citep{slingo1991,stephens1978,stephens1984}, large-scale precipitation, boundary fluxes of sensible and latent heat and of momentum, long-wave and short-wave radiation \citep{sasamori1968,lacis2018}, vertical and horizontal diffusion \citep{laursen1989,louis1979,louis1981} and a land surface scheme with five diffusive layers for temperature and a bucket model for soil hydrology. 

The \acrshort{lsg} is a three-dimensional global ocean general circulation model based on primitive equations under the assumption of large spatial and temporal scales, using the Boussinesq and hydrostatic approximations and neglecting vertical friction \citep{maier1993,maier1992}. Turbulent motions are parameterized by a vertical oceanic diffusion coefficient. The \acrshort{lsg} model is used to calculate the heat flux from the deep ocean to the mixed-layer due to advective and convective processes, i.e., advection, horizontal diffusion, vertical transport, vertical diffusion, convective adjustments. In contrast, the mixed-layer ocean model is used to compute the temperature tendencies of the mixed-layer due to heat exchanges with the atmosphere. 

The thermodynamic sea ice model is based on the zero-layer model of \cite{semtner1976}. This model computes the evolution of the sea ice thickness and surface temperature from the energy balances at the top and bottom of a sea ice-snow layer. The sea ice-snow layer is assumed to have a linear temperature gradient and to have no capacity to store heat. The sea ice concentration is binary, i.e., either a grid cell is fully sea ice covered or open water. 

We run the \acrshort{plasim-lsg} simulations at statistically stationary state with a fixed pre-industrial effective CO\textsubscript{2} volume mixing ratio of 280 ppmv. We perform the runs without diurnal cycle and with seasonal cycle of the solar radiation and each year has 360 days. For the atmosphere, we use a horizontal spectral resolution of T21 (triangular truncation at wavenumber 21 $\sim$ 5.625\degree\ x 5.625\degree\ on the corresponding Gaussian grid), ten non-equally spaced sigma-levels up to about 40 hPa in the vertical, a computational time step of 45 minutes and an output time step of one day. The computational and output time steps of the mixed-layer ocean and sea ice models are one day. The \acrshort{lsg} is run on a 2.5\degree\ x 5\degree\ staggered E-type grid \citep{arakawa1977} in the horizontal, with 22 levels in the vertical and with computational and output time steps of five days.

\subsection{Pan-Arctic sea ice area in the control run}\label{subsec2}
We consider a control run (CTRL) of 3000 years at equilibrium state (Figure 1; labeled as model years 501-3500; model years 1-500 are discarded as spin-up). From the control run, we derive several experiments with the rare event algorithm (see section 2.3.2). We consider the statistics of the pan-Arctic sea ice area 
\begin{equation}
    A (t) = \sum_{\phi \ge \phi_{min}} \sum_{\lambda} SIC_{\phi,\lambda} (t) \cdot G_{\phi,\lambda},
\end{equation}      
where $SIC_{\phi,\lambda}(t)$ is the sea ice concentration at time $t$ in a grid cell centered at latitude $\phi$ and longitude $\lambda$, $G_{\phi,\lambda} =  \int_{\phi-\frac{\Delta \phi}{2}}^{\phi+\frac{\Delta \phi}{2}} \int_{\lambda-\frac{\Delta \lambda}{2}}^{\lambda+\frac{\Delta \lambda}{2}} R^2 cos(\phi^{\prime}) d\lambda^{\prime} d\phi^{\prime}$ is the grid cell area, $R$ is the earth radius, $\Delta\phi$ and $\Delta \lambda$ are the angular distances between two grid points in the meridional and zonal direction. The summation in (1) includes all ocean grid boxes north of 40\degree N  (i.e. $\phi_{min} \cdot \frac{180\degree}{\pi} = 40\degree$; the black circle in Figure 1(a) shows the southern boundary of the domain over which the pan-Arctic sea ice area is computed). The land-sea mask is binary (i.e. a grid cell is either completely ocean or land).

\begin{figure*}[!htb]
    \vspace{0.6ex}
    \center{\includegraphics[width=\textwidth,trim={0.03cm 0.03cm 0.03cm 0.03cm},clip]
    {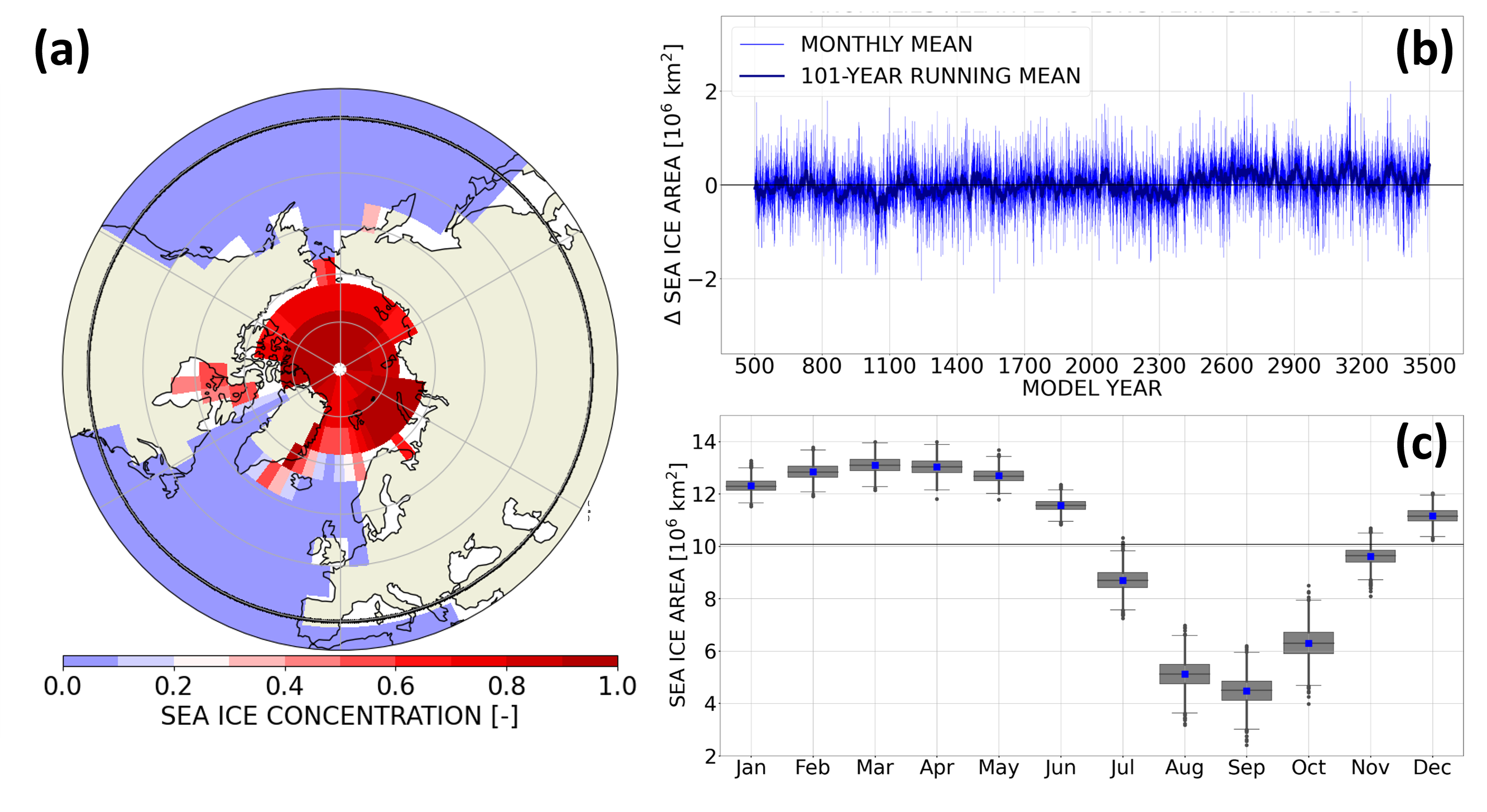}}
    \vspace*{-7mm}
    \caption{\label{} \acrshort{plasim-lsg} control run (model years 501-3500): (a) Black circle shows the southern boundary of the domain over which the pan-Arctic sea ice area is computed. Shading shows the climatological annual mean sea ice concentration [-] field. (b) Time series of monthly mean pan-Arctic sea ice area anomalies [10\textsuperscript{6} km\textsuperscript{2}] relative to the climatology of the control run for (blue) the original monthly values and (darkblue) the 101-year running mean. (c) Distributions of monthly mean pan-Arctic sea ice area [10\textsuperscript{6} km\textsuperscript{2}] with respect to the 3000 model years. The averages and medians are given by the blue squares and the horizontal lines in the boxes. The boxes denote interquartile ranges and the maximum whisker length is defined as 1.5 times the interquartile range. The horizontal gray line shows the annual mean pan-Arctic sea ice area.}
\end{figure*} 

The monthly mean pan-Arctic sea ice area anomalies show interannual variability superimposed on weak fluctuations on multi-year to centennial time scales (Figure 1(b)). For example, the sea ice area tends towards positive anomalies between model years 2500 and 2700 and towards negative anomalies around model year 1000 (Figure 1(b)). The annual average, the amplitude of the seasonal cycle and the timing of the annual minimum and maximum of the pan-Arctic sea ice area are representative of the observed Arctic sea ice climatology between 1979 and 2015 (cf. Figures 1(a,c) and Supplementary Figure S1(a,b); we used data from \cite{eumetsat1,eumetsat2} to compare the representation of sea ice in \acrshort{plasim-lsg} with observations (Supplementary Figure S1)). Nevertheless, compared to observations, \acrshort{plasim-lsg} has a delayed melting period, a positive \newacronym{sic}{SIC}{sea ice concentration} \acrshort{sic} bias from the Greenland to the Kara Sea and a negative \acrshort{sic} bias western of Greenland (cf. Figures 1(a,c) and Supplementary Figure S1(a,b)). Despite this, the representation of sea ice dynamics and statistics in \acrshort{plasim-lsg} is sufficient for the purpose of this study.
 
We study extreme negative anomalies of \newacronym{febsep}{FEBSEP}{February-September} \acrfull{febsep} averaged pan-Arctic sea ice area. We classify a sea ice area anomaly $I^{\prime}(t)$ as extremely negative if $I^{\prime}(t) \le -n \sigma_{CTRL}$, where $\sigma_{CTRL}$ is the standard deviation of the control run and $n$ is a real-valued number being defined in section 3. 

\subsection{Rare event algorithm}\label{sec3}

\subsubsection{Description of the method}\label{subsec1}
In climate modeling and weather forecast, ensemble simulations provide a distribution of possible trajectories of the state of the atmosphere or of the climate system. In a conventional ensemble simulation, most of these trajectories correspond to typical fluctuations of an observable, e.g. the pan-Arctic sea ice area. Only a few trajectories lead to extreme states of the observable, which hampers robust statistical and dynamical studies on events that correspond to the tail of the distribution. The genealogical selection algorithm presented in \cite{giardina2011simulating,ragone2018,ragone2019,ragone2021} is designed to guide ensemble simulations to oversample rare dynamical trajectories that lead to extreme anomalies of the time-average of an observable where the averaging time is longer than the typical decorrelation time of the observable. This algorithm is therefore well suited to study the statistics of the seasonally averaged pan-Arctic sea ice area.

We describe the main steps of the algorithm in the following and refer to \cite{ragone2018} and \cite{ragone2019,ragone2021} for more details. We denote $X(t)$ the vector of all model variables at time $t$. We consider an ensemble of $N$ trajectories $\lbrace X_n(t) \rbrace$ $(n = 1,2,...,N)$, an observable $A(\lbrace X_n(t) \rbrace)$ (i.e., a function that maps for each trajectory the state vector to a single scalar), a total simulation time $T_a$ and a resampling time $\tau_r$. We start the ensemble simulation from statistically independent initial conditions. We perform at regular times $t_i = i \cdot \tau_r\ (i = 1,2,...,\frac{T_a}{\tau_r})$ a resampling procedure in which trajectories are killed or generate a random number of replicates depending on weights that are related to the magnitude of the time-average of the observable during the past interval. The weights are defined as

\begin{equation}
    w_n^i = \frac{e^{k \int_{t_{i-1}}^{t_{i}} A(\lbrace X_n(t) \rbrace) \,dt\ }}{R_i} \text{        with       } R_i = \frac{1}{N} \sum_{n=1}^{N} e^{k \int_{t_{i-1}}^{t_{i}} A(\lbrace X_n(t) \rbrace) \,dt\ }, 
\end{equation}


\vspace{3mm}

\noindent where $R_i$ is a normalization term and $k$ is a biasing parameter. With a positive (negative) $k$, the weights favour the replication of trajectories leading to large (small) values of the time-average of the observable during the past interval, while trajectories leading to small (large) values of the time-average of the observable are likely to be killed. The absolute value of $k$ controls the strength of the selection. As described in \cite{ragone2018}, each resampling is performed in such a way that the ensemble size remains constant and equal to $N$. After the resampling, we slightly perturb the surface pressure field of each trajectory to enable copies of the same trajectory to separate from each other during the subsequent simulation interval (the perturbation is performed as described in the Supporting Information of \cite{ragone2018}). When the final time $T_a$ is reached, we perform one last resampling and reconstruct an effective ensemble by attaching at each resampling event from the end to the beginning of the simulation the ancestors to the pieces of surviving trajectories. All trajectories that did not survive until the end of the simulation are discarded.

As described in \cite{ragone2018} and \cite{ragone2019}, one obtains for a large ensemble size $N$ the importance sampling formula

\addtolength{\jot}{0.5em}
\begin{equation}
    \mathbb{P}_k(\lbrace X_n(t) \rbrace_{0 \leq t \leq T_a}) \underset{N\to\infty}{\sim} \frac{e^{k \int_{0}^{T_a} A(\lbrace X_n(t) \rbrace) \,dt\ }}{\mathbb{E}_0[e^{k \int_{0}^{T_a} A(\lbrace X_n(t) \rbrace) \,dt\ }]} \mathbb{P}_0(\lbrace X_n(t) \rbrace_{0 \leq t \leq T_a}),
\end{equation}

\vspace{3mm}

\noindent where $\mathbb{P}_0$ and $\mathbb{P}_k$ are the probability densities of trajectories in the standard ensemble simulation and in an ensemble generated with the algorithm and $\mathbb{E}_0$ is the expectation value with respect to $\mathbb{P}_0$. The importance sampling formula describes the ratio of probability densities of trajectories in a simulation obtained with the rare event algorithm to the probability density of trajectories according to the real model statistics. Consequently, expectation values (e.g. composites, return times) with respect to the real model statistics can be computed from the data obtained with the rare event algorithm by weighting the contribution of each trajectory to sample averages by the inverse of the exponential factor in (3) (see \citeauthor{ragone2018},~\citeyear{ragone2018}; \citeauthor{ragone2019},~\citeyear{ragone2019}; Supporting Information of \citeauthor{ragone2021},~\citeyear{ragone2021} for more details).

\subsubsection{Set-up of the rare event algorithm experiments}\label{subsec2}

We use as target observable $A(\lbrace X_n(t) \rbrace)$ the pan-Arctic sea ice area and perform two sets of experiments with the rare event algorithm. One set is performed with a resampling time of $\tau_r = 30$ days, the other with a resampling time of $\tau_r = 5$ days. Both resampling times are chosen to be smaller than the decorrelation time of the pan-Arctic sea ice area (the autocorrelation function of sea ice area anomalies drops  for the first time below 1/e after about 75 days). In this way, the selection favours the survival of trajectories characterized by large negative time-persistent sea ice area anomalies during the interval prior to a resampling event. In contrast to the resampling time of $\tau_r$ = 30 days, the resampling time of $\tau_r = 5$ days is not larger than the persistence time scale of the large-scale atmospheric circulation \citep{baldwin2003} and is only slightly larger than the typical persistence of synoptic-scale atmospheric fluctuations \citep{van1957power}. We therefore expect that the experiments with a resampling time of $\tau_r = 5$ days efficiently sample extremely low sea ice states both due to anomalies in the oceanic thermal state and due to anomalies in the atmospheric circulation that are in the order of or larger than the upper range of the synoptic time scale. In contrast, the experiments with a resampling time of $\tau_r$ = 30 days are primarily designed to increase the sampling efficiency of oceanic drivers of low sea ice states. In this case, we expect atmospheric drivers of sea ice reduction to be sampled only if they are characterized by an anomalously large persistence (e.g. an one month long persistent negative phase of the summer \acrshort{ao}).

Both sets of experiments include $M=5$ ensemble simulations labelled with $m=1,...,M$ (Table 1). Each ensemble has $N=600$ trajectories and a total simulation time of $T_a = 240$ days between February 1st and September 30th. We take different values of the biasing parameter for the different experiments: $k = -0.06\ \cdot $ 10\textsuperscript{-6} km\textsuperscript{-2} day\textsuperscript{-1} for ensemble $m = 1$, $k = -0.05\ \cdot $ 10\textsuperscript{-6} km\textsuperscript{-2} day\textsuperscript{-1} for ensembles $m = 2$ and $m = 4$ and $k = -0.04\ \cdot $ 10\textsuperscript{-6} km\textsuperscript{-2} day\textsuperscript{-1} for ensembles $m=3$ and $m=5$. While the precise values of $k$ is chosen empirically, a reasonable order of magnitude of the parameter can be derived using a scaling argument presented in \cite{ragone2019} and in the Supporting Information of \cite{ragone2021}. According to this scaling argument, the selected $k$-values correspond to a shift of the peak of the distribution of February-September mean pan-Arctic sea ice area anomalies to values between $-0.5\ \cdot $ 10\textsuperscript{6} km\textsuperscript{2} and $-0.75\ \cdot $ 10\textsuperscript{6} km\textsuperscript{2}, which corresponds to a range of estimated return times between 100 and 1000 years (see section 3.1).

For each ensemble, the inital condition for trajectory $n = 1,...,N$ is taken from February 1st in year $500+m+5(n-1)$ of the control run (Table 1). In order to have a baseline for the statistics, we also subdivide the 3000-year control run into five 600-member control ensembles such that the rare event algorithm ensemble $m$ has exactly the same initial conditions as the control ensemble $m$. Sampling the initial conditions with a gap of five years ensures that sea ice area anomalies in the different trajectories within an ensemble are approximately independent from each other.

\begin{table*}[ht]
\caption{600-member rare event algorithm ensemble simulations running between February 1st and September 30th. This set-up applies both to the experiments with a resampling time of $\tau_r$ = 5 days and to the experiments with a resampling time of $\tau_r$ = 30 days.} 
\vspace{-0.5em}
\centering 
\begin{tabular}{| c | c | c |} 
\hline 
$m$ & Model years for initial condition & Biasing parameter $k$ [10\textsuperscript{-6} km\textsuperscript{-2} day\textsuperscript{-1}] \\ [0ex] 
\hline\hline 
1 & 501, 506,..., 3496 & -0.06 \\ 
\hline
2 & 502, 507,..., 3497 & -0.05 \\
\hline
3 & 503, 508,..., 3498 & -0.04 \\
\hline
4 & 504, 509,..., 3499 & -0.05 \\
\hline
5 & 505, 510,..., 3500 & -0.04 \\ [0ex] 
\hline 
\end{tabular}
\label{table:nonlin} 
\end{table*}

\section{Results}\label{sec4}

\subsection{Importance sampling of summer seasons with extremely low pan-Arctic sea ice area}\label{subsec1}

Our goal of using the rare event algorithm is to improve the sampling efficiency of extreme negative pan-Arctic sea ice area anomalies on average over the extended summer season between February and September. We show the seasonal evolution of daily pan-Arctic sea ice area anomalies merged over the two rare event algorithm experiments with $k=-$0.05 $\cdot$  10\textsuperscript{-6} km\textsuperscript{-2} day\textsuperscript{-1} (Figure 2(a,b)). Throughout the season, trajectories generated with the rare event algorithm show a systematic shift towards lower sea ice area values compared to the control run. The amplitude of this shift does not differ between the 5-days and the 30-days resampling time experiments and is more than twice as large after July than between February and June. The amplification of the shift during July coincides with an increase in the climatological sea ice area variability of the control run (cf. Figures 2(a,b) and Figure 1(c)). 

The systematic biasing towards negative sea ice area anomalies both in the 5-days and 30-days resampling time experiments emerges from the trajectory resampling at intervals shorter than the decorrelation time of the sea ice area. During each resampling event, trajectories characterized by large negative time-persistent sea ice area anomalies form replicas at the expense of trajectories with less negative sea ice area anomalies. After the simulation, the surviving pieces of trajectories are reconstructed into an effective ensemble. As a consequence, the rare event algorithm efficiently performs importance sampling of trajectories that lead to extreme negative February-September time-averaged sea ice area anomalies (Figure 2). The control distribution of February-September mean sea ice area anomalies has a standard deviation of about 0.25 $\cdot$ 10\textsuperscript{6} km\textsuperscript{2} and a minimum value of $-0.80$ $\cdot$ 10\textsuperscript{6} km\textsuperscript{2}. In contrast, the sea ice area anomalies obtained with the rare event algorithm fluctuate around a mean value of $-0.6\ \cdot$ 10\textsuperscript{6} km\textsuperscript{2} and have minima at $-0.89\ \cdot$ 10\textsuperscript{6} km\textsuperscript{2} (30-days resampling time experiments) and at $-0.99\ \cdot$ 10\textsuperscript{6} km\textsuperscript{2} (5-days resampling time experiments). The simulations with the rare event algorithm require the same computational cost as the control run, but strongly populate the lower tail of the distribution of \acrshort{febsep} mean pan-Arctic sea ice area anomalies (Figure 2(c,d); we refer to the Supplementary Information S2 and Supplementary Figures S5 and S6 for more details about the adequacy of the rare event simulations used in this work, i.e., of the extent to which the effective ensembles contain sufficiently enough different initial-time ancestors with a non-zero number of final-time decendants to gather robust statistics).

\begin{figure*}[!htp]
    \vspace{0.6ex}
    \center{\includegraphics[width=\textwidth,trim={0.03cm 0.03cm 0.03cm 0.03cm},clip]
    {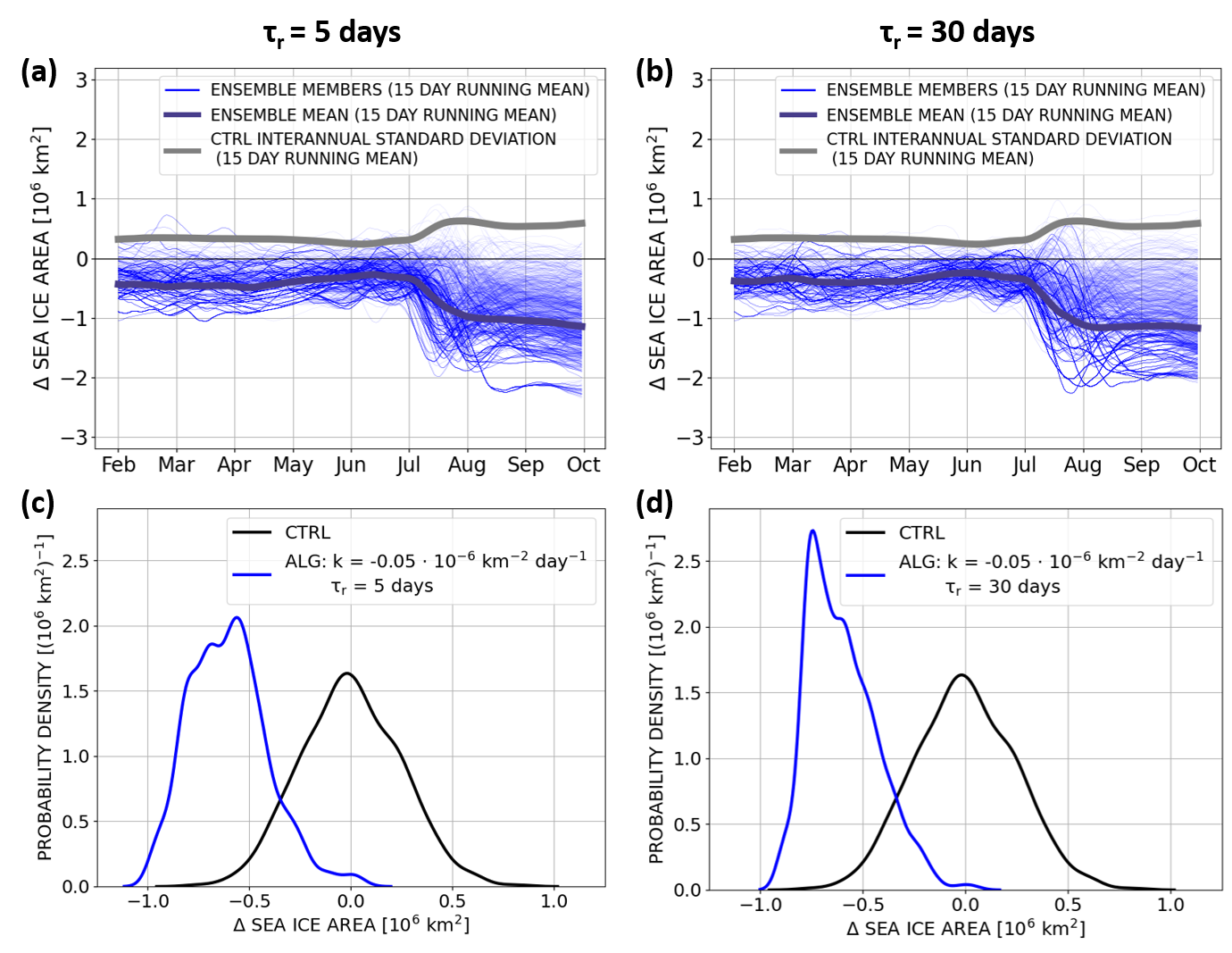}}
    \vspace*{-7mm}
    \caption{\label{} (a-b) Trajectories (thin blue lines) and ensemble mean (thick blue line) of daily pan-Arctic sea ice area anomalies in rare event algorithm ensemble simulations two and four, i.e. with $k=-0.05\ \cdot $ 10\textsuperscript{-6} km\textsuperscript{-2} day\textsuperscript{-1}, for (a) a resampling time of 5 days and (b) a resampling time of 30 days. The anomalies are evaluated relative to the daily climatology of the control ensembles two and four. Only the trajectories that survived until the final simulation time are plotted and are used for the computation of the ensemble mean. The gray lines show the climatological standard deviation of daily sea ice area anomalies in the control ensembles m=2 and m=4. (c-d) Probability distribution functions of February-September mean pan-Arctic sea ice area anomalies relative to the control climatology for (c) a resampling time of 5 days and (d) a resampling time of 30 days. (Black) Anomalies of the control run itself merged over ensembles m=2 and m=4 and (blue) anomalies in the rare event algorithm experiments corresponding to ensembles m=2 and m=4, i.e. with $k=-0.05\ \cdot\ $10\textsuperscript{-6} km\textsuperscript{-2} day\textsuperscript{-1} (exclusively based on the trajectories that survived until the end of the simulation).}
\end{figure*} 
 
Return times are an important characterization of extreme events. They indicate the average waiting time between events of size larger than a given amplitude. We compare return times of \acrshort{febsep} mean pan-Arctic sea ice area anomalies between the control run and the experiments with the rare event algorithm (Figure 3(a,b)). For the computation of the return times, we proceed similarly to the modified block maximum estimator presented in \cite{lestang2018} and in the Supporting Information of \cite{ragone2018}. Accordingly, an estimate of the return time is given by 
\begin{equation}
    r(a) = -\frac{\Delta T}{\ln[1-\frac{1}{N} \sum_{n=1}^{N} \mathbbm{1}_a(I_n)]} \text{        with       } \mathbbm{1}_a(I_n) = \left\{\begin{array}{ll} 1, & I_n \leq a \\ 0, & I_n > a  \end{array}\right. ,
\end{equation}  
where $I_n$ is the February-September time-averaged pan-Arctic sea ice area anomaly in trajectory $n$, $a$ is the return level of a sea ice area anomaly, $N$ the number of trajectories, $\Delta T$ is the block length of one year and $\mathbbm{1}_a(I_n)$ is the indicator function. As described in \cite{lestang2018} and in the Supporting Information of \cite{ragone2018}, equation (3) can be used to compute the return times from the simulations with the rare event algorithm. Note that for large values of the return time, the formula in (4) is approximately equivalent to $r(a) = T_d/\sum_{n=1}^{N} \mathbbm{1}_a(I_n)$, where $T_d = N\Delta T$ is the total length of the time series (see \cite{lestang2018}). 

In Figure 3, the red curves show return time estimates obtained from the 3000-year control run. In order to estimate uncertainty ranges, we subdivide the control run into five 600-member ensembles (see section 2.3.2) providing five return time estimates. We compute the average over these five estimates (black curve) and use their empirical standard deviation to construct 95\% confidence intervals (black shading; see Supplementary Information S1 for more details). In the same way, the blue lines and blue shading show the average return time estimates and confidence intervals obtained from the overlapping output of at least three out of the five 600-member rare event algorithm experiments (note that rare event algorithm experiments with slightly different $k$ values cover slightly different ranges of sea ice area anomalies). Both for the 5-days and for the 30-days resampling time experiments, the return curves for the control run and for the rare event algorithm overlap in a range of sea ice area anomaly values. This range corresponds to an overlap between the probability distribution functions of sea ice area anomalies obtained with the control and rare event algorithm ensembles (cf. Figures 3(a), 2(c), Supplementary Figure S2(a,c) and Figures 3(b), 2(d), Supplementary Figure S2(b,d)). Consequently, the rare event algorithm consistently computes the probabilities that trajectories generated with the algorithm have with respect to the real model statistics (an analysis of the differences between the control and rare event algorithm estimates of the return levels shows that both estimates are statistically consistent with each other at least for a return time range between a couple of decades and about 750 years (Supplementary Figure S3(a,b))).   

The major advantage of computing the return times with the rare event algorithm compared to the control run is the access to much rarer events with the former than with the latter. From the control run, we cannot accurately estimate return times of more than 1000 years. The rare event algorithm experiments are conducted with the same computational cost as the control run (i. e. 3000 years), but allow to compute return times up to $10^5$ years with uncertainty ranges comparable with the control run ones for return times of several hundreds of years. Hence the algorithm increases the sampling efficiency of extreme negative February-September mean sea ice area anomalies by two to three orders of magnitude. The plateaux at return times larger than 10\textsuperscript{5} years (Figure 4(a)) and larger than 5 $\cdot$ 10\textsuperscript{4} years (Figure 4(b)) are due to undersampling as discussed in \cite{lestang2018} and in the Supporting Information of \cite{ragone2018}.    

The result has the following implications. Firstly, the rare event algorithm provides access to ultra-rare sea ice area anomalies that do not occur in the control run. An unrealistic amount of computational cost would be required to observe them with direct sampling. Secondly, the algorithm enriches the general statistics of the extremes. While the control run delivers only a few events with a sea ice area anomaly in the order of three standard deviations away from the climatology, hundreds of them are available in the experiments with the algorithm. Thirdly, the return curve obtained with the 5-days resampling time experiments extends only to marginally larger amplitudes of sea ice area anomaly values than the one obtained with the 30-days resampling time experiments. Thus, the targeted sampling of drivers of sea ice reduction acting on the upper range of synoptic and on submonthly time scales only marginally increases the magnitude of the most extreme negative February-September mean pan-Arctic sea ice area anomalies. We will further address this point in section 4. 
 
\begin{figure*}[!htp]
    \vspace{0.6ex}
    \center{\includegraphics[width=\textwidth,trim={0.03cm 0.03cm 0.03cm 0.03cm},clip]
    {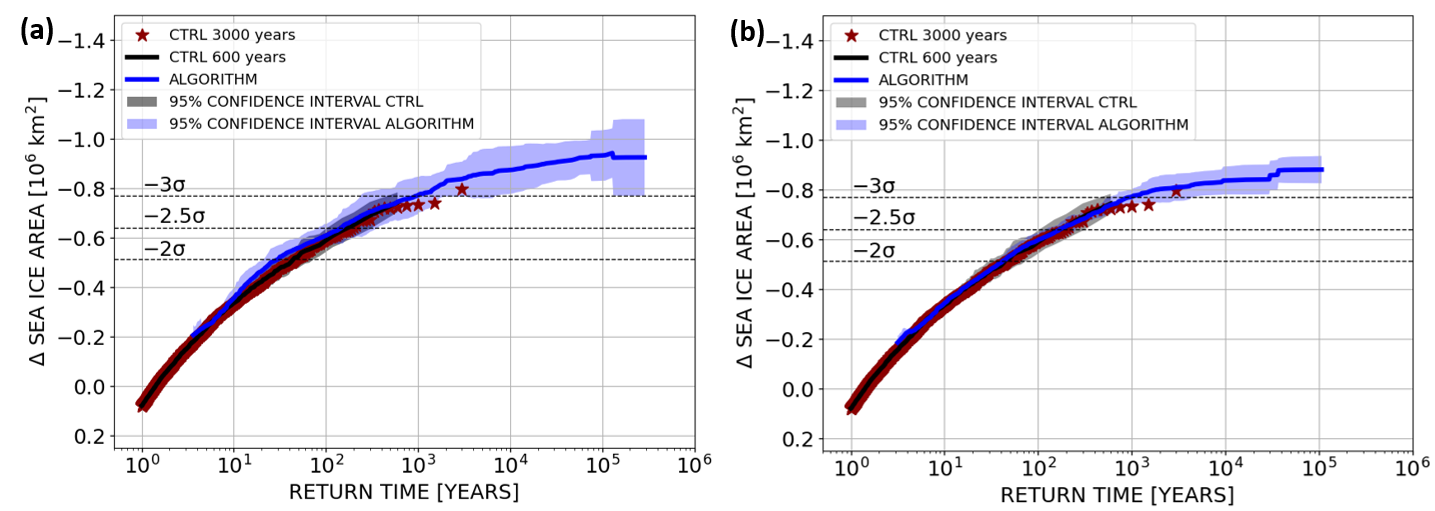}}
    \vspace*{-5mm}
    \caption{\label{} Return curves for February-September mean pan-Arctic sea ice area anomalies relative to the control run. (Red stars) The direct estimate of return times from the 3000 year control run, (black line) the average estimate over the five 600-member ensembles of the control run and (blue line) the average estimate over the overlap of at least three out of five 600-member rare event algorithm experiments. Shading denotes the 95\% confidence interval obtained from the statistics of the three to five estimates assuming a student's t-distributed estimator (see Supplementary Information S1). (a) Rare event algorithm with 5-days resampling time and (b) with 30-days resampling time. The dashed lines in (a) and (b) indicate anomalies of minus two, minus two and a half and minus three standard deviations with respect to the control run.}
\end{figure*}

\subsection{Sea ice conditions and state of the atmosphere during summer seasons of extremely low sea ice area}\label{subsec2}

We perform composite analyses to investigate and characterize the average states of the sea ice and of the atmosphere during summer seasons of extreme negative February-September mean pan-Arctic sea ice area anomalies. Composites mathematically correspond to conditional expectation values that we estimate as 

\addtolength{\jot}{0.5em}
\begin{equation}
    \mathbb{E}_0[O(\lbrace X_n(t) \rbrace)\ |\ I_n \leq -m\sigma] = \frac{\mathbb{E}_0[O(\lbrace X_n(t) \rbrace) \cdot \mathbbm{1}(-I_n-m\sigma)]}{\mathbb{E}_0[\mathbbm{1}(-I_n-m\sigma)]}, 
\end{equation} 

\vspace{2mm}

\noindent where $O(\lbrace X_n(t) \rbrace)$ and $I_n$ are the anomalies of a target climate variable (e.g. 500 hPa geopotential height) and of the February-September mean pan-Arctic sea ice area in trajectory $n$, $\sigma$ is the control run standard deviation of $I$, $m$ a positive integer and $\mathbbm{1}(-I_n-m\sigma)$ is the indicator function with
\begin{equation}
\mathbbm{1}(-I_n-m\sigma) = \left\{\begin{array}{ll} 1, & -I_n-m\sigma \geq 0 \\
         0, &  -I_n-m\sigma < 0  \end{array}\right..
\end{equation}

\noindent All anomalies are evaluated relative to the climatology of the control ensembles and $\mathbb{E}_0$ is the expectation operator with respect to the unbiased model statistics (i.e. the statistics of the control run). By using equation (3), composites can be computed from the output of the rare event algorithm as described in \cite{ragone2019} and in the Supporting Information of \cite{ragone2018} and of \cite{ragone2021}.
  
One advantage of the rare event algorithm is that it provides more accurate composite estimates for large return times than the control run. For the subsequent analyses, we compute composite estimates for each of the five rare event algorithm ensembles and consider the average over the five estimates respectively. We assess the statistical significance of the composite maps by applying a two-sided t-test to the set of the five estimates (see Supplementary Information S1). We apply the same procedure to five 600-member control ensembles obtained from the 3000-year control run (see section 2.3.2). All subsequent results obtained with the rare event algorithm are based on the experiments with a resampling time of $\tau_r = 5$ days. As explained in section 2.3.2, this value of resampling time allows to increase the sampling efficiency of a broader range of physical processes compared to a resampling time of $\tau_r = 30$ days.  

\subsubsection{Seasonal mean states of the sea ice and of the atmosphere}\label{subsubsec1}
We compare the control run and rare event algorithm estimates of average seasonal mean \acrshort{sic}, \newacronym{t2m}{T2M}{two metre temperature} \acrfull{t2m} and \newacronym{z500}{Z500}{500 hPa geopotential height} \acrfull{z500} anomalies during summer seasons of extreme negative pan-Arctic sea ice area anomalies with return times of more than 200 years (i.e. $I_n \leq -2.5\sigma$) (Figure 4). For this threshold, we have access to several hundreds of events with the algorithm and to 13 events with the control run. While the algorithm and control estimates consistently indicate a warm Arctic state during extremely low sea ice seasons, the benefit of the algorithm manifests in the larger statistical accuracy of the estimates compared to the control run. The algorithm composites indicate statistically significant negative mean \acrshort{sic} anomalies in the North Atlantic side of the Arctic, along the Russian coast and in the marginal seas of the Pacific side of the Arctic Ocean (Figure 4(d)). Statistically significant positive \acrshort{t2m} anomalies occur over the entire Arctic Ocean with largest amplitudes over the Canadian Archipelago, Greenland and Kara Seas (Figure 4(e)). The anomalously warm surface conditions are accompanied by positive \acrshort{z500} anomalies over the Arctic reaching statistical significance from the Canadian Archipelago over Svalbard to the Barents and Kara Seas. Compared to the algorithm, the control run estimates of the mean \acrshort{sic}, \acrshort{t2m} and \acrshort{z500} anomalies are statistically significant in a much smaller amount of grid cells. Apart from the formal statistical significance, the larger amount of extreme events available through the algorithm decreases the likelihood that the composite maps include noisy patterns that are not statistically related to the sea ice area anomalies and that are occurring in the control run by the effect of sampling (Figure 4(b,c,e,f)).
      
\begin{figure*}[!htp] 
    \vspace{0.6ex}
    \center{\includegraphics[width=\textwidth,trim={0.03cm 0.03cm 0.03cm 0.03cm},clip]
    {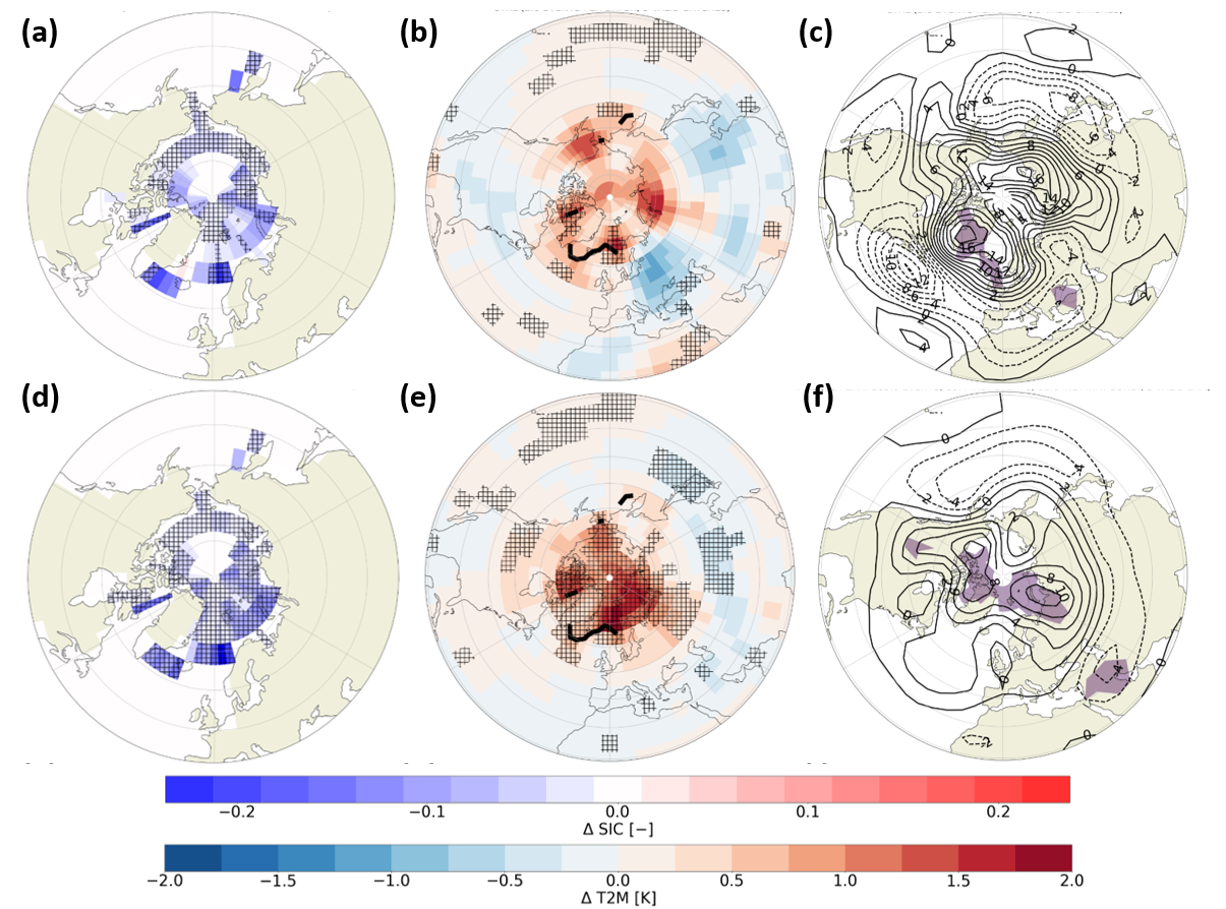}}
    \vspace*{-5mm}
    \caption{\label{} Composite mean February-September averaged (a,d) sea ice concentration (SIC [-]), (b,e) two metre temperature (T2M [K]) and (c,f) 500 hPa geopotential height (Z500 [gpm]; contour interval two gpm) anomalies conditional on extreme negative February-September mean pan-Arctic sea ice area anomalies equal or smaller than -2.5 standard deviations of the control ensembles (corresponds to extremes with return times of more than 200 years). Composite maps are presented as an average over (a-c) the five control and (d-f) the five rare event algorithm experiments with a resampling time of $\tau_r = 5$ days. All anomalies are estimated with respect to the climatology of the control ensembles. The hatching in (a,b,d,e) and shading in (c,f) denote statistical significance at the 5\% level according to a two-sided t-test (see the Supplementary Information S1 for more details). The black contour line in (b,e) indicates the climatological February-September mean sea ice edge defined as the 15\% sea ice concentration line.}
\end{figure*} 

\subsubsection{Seasonal evolution of the states of the sea ice and of the atmosphere}\label{subsubsec2}
We are interested in the physical processes favouring anomalously warm Arctic states with extremely low pan-Arctic sea ice area during summer. From the physical point of view, understanding the drivers of extremely low pan-Arctic sea ice area is hindered by at least two factors. Firstly, a variety of feedback mechanisms occur in the Arctic climate system \citep[e.g.][]{screen2010,serreze2006} and can lead to ambiguities of cause and effect relationships between different quantities. This applies, for example, to the relationship between the anomalously warm Arctic atmosphere and the negative sea ice concentration anomalies shown in Figure 4. The negative sea ice concentration anomalies are potentially driven by the anomalous warm atmosphere. At the same time, however, the positive \acrshort{t2m} and \acrshort{z500} anomalies are potentially a consequence of the anomalous heat flux from the ocean to the atmosphere that results from the reduced sea ice cover. Secondly, extremely low summer pan-Arctic sea ice area may be generated in two different ways. Are we sampling trajectories in which processes on intra-seasonal time scales lead to an extreme reduction of the pan-Arctic sea ice area within a season? Or are we sampling trajectories in which extreme negative February-September mean pan-Arctic sea ice area anomalies are related to preconditioning \citep[cf.][]{chevallier2012role,holland2011inherent,kauker2009adjoint}?

We firstly identify to what extent extreme negative February-September mean pan-Arctic sea ice area anomalies are related to intra-seasonal sea ice reduction vs. pre-existing anomalies in the sea ice state originating from the previous winter. For this purpose, we analyse the seasonal evolution of mean pan-Arctic sea ice area and sea ice volume anomalies conditional on the occurrence of summer seasons of extremely low pan-Arctic sea ice area (Figure 5). On average, extreme negative \acrshort{febsep} pan-Arctic sea ice area anomalies are related both to persistent sea ice area anomalies originating from the previous winter and to anomalous sea ice area reduction between late spring and late summer (Figure 5(a)). The sea ice area anomalies between February-March and May-June are in the order of one and two-thirds of the interannual standard deviation and have an approximately constant level of about $-0.2\ \cdot $ 10\textsuperscript{6}\ km\textsuperscript{2} to $-0.5\ \cdot $ 10\textsuperscript{6}\ km\textsuperscript{2}. In contrast, the sea ice area anomalies in August-September are in the order of $-1\ \cdot $ 10\textsuperscript{6} km\textsuperscript{2} to $-1.5\ \cdot $ 10\textsuperscript{6}\ km\textsuperscript{2}, two to three times as large in magnitude as the interannual standard deviation. As a consequence, the anomalies in the sea ice area reduction between February-March and August-September are about twice as large in magnitude as the pre-existing sea ice area anomalies in late winter. The relative importance of intra-seasonal sea ice area reduction over pre-existing sea ice area anomalies slightly increases with the amplitude of the extremes of the February-September mean pan-Arctic sea ice area anomalies (Figure 5(a)). 

In contrast to the pan-Arctic sea ice area, pan-Arctic sea ice volume anomalies are directly related to the amount of energy required to produce these anomalies and their seasonal evolution does not directly depend on the open-water-formation-efficiency given by the sea ice thickness field. Pre-existing pan-Arctic sea ice volume anomalies originating from the previous winter are three to four times as large as the anomalies in their reduction from February-March to August-September (Figure 5(b)). This result points at a strong contribution of preconditioning of extreme anomalies of the \acrshort{febsep} mean pan-Arctic sea ice area through an extremely low winter pan-Arctic sea ice volume, which is both due to negative anomalies in the winter pan-Arctic sea ice area and in the winter pan-Arctic mean sea ice thickness field (cf. Figure 5 and Supplementary Figure S4(a,b)). On top of the preconditioning, an anomalous intra-seasonal reduction in the sea ice volume occurs between April-May and July-August. This suggests a contribution of intra-seasonal dynamics to extreme negative \acrshort{febsep} sea ice area anomalies in addition to the preconditioning.
  
\begin{figure*}[!htb]
    \vspace{0.6ex}
    \center{\includegraphics[width=\textwidth,trim={0.03cm 0.03cm 0.03cm 0.03cm},clip]
    {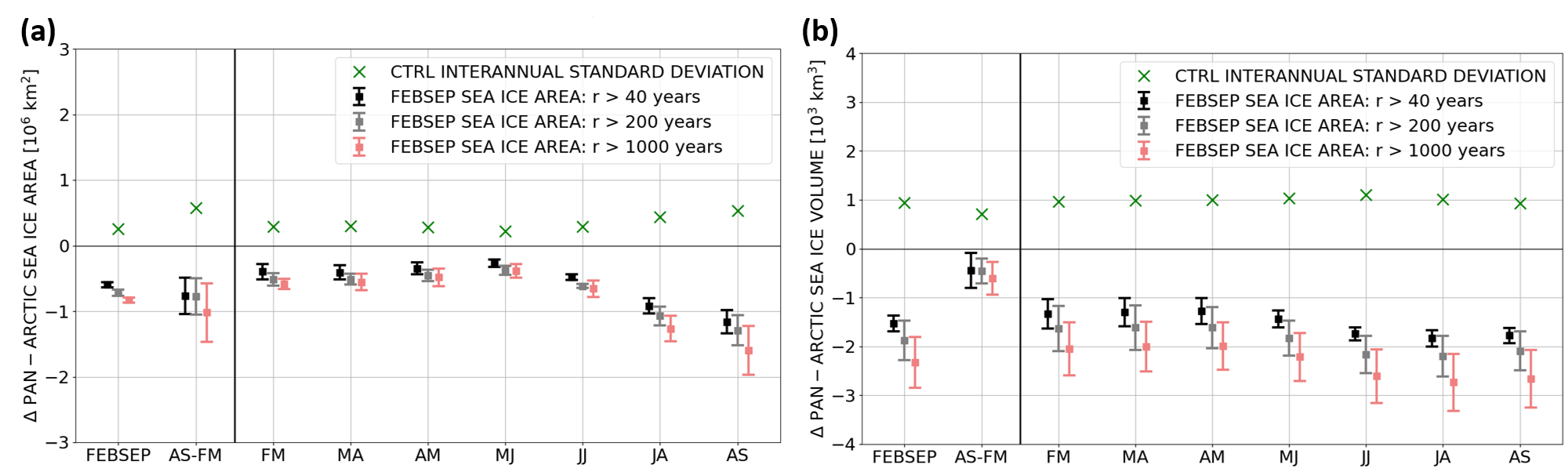}}
    \vspace*{-5mm}
    \caption{\label{} Rare event algorithm experiments with a resampling time of $\tau_r = 5$ days: Composite mean (a) pan-Arctic sea ice area anomalies [$10^{6}$ km\textsuperscript{2}] and (b) pan-Arctic sea ice volume anomalies [$10^{3}$ km\textsuperscript{3}] conditional on extreme negative February-September mean pan-Arctic sea ice area anomalies equal or smaller than (black) -2 standard deviations, (gray) -2.5 standard deviations and (red) -3 standard deviations of the control ensembles (roughly corresponds to extremes with return times of more than 40, 200 and 1000 years). The anomalies are computed with respect to the control climatologies of the five control ensembles. The composite estimates are presented as an average over the five rare event algorithm ensembles and the error bars represent 95\% confidence intervals (see Supplementary Information S1). "AS-FM" denotes the difference between \newacronym{as}{AS}{August-September} \acrfull{as} and \newacronym{fm}{FM}{February-March} \acrfull{fm} and the green "x" markers indicate the climatological standard deviation in the control run.}
\end{figure*} 

We analyse the seasonal evolution of \acrshort{t2m} and \acrshort{z500} anomalies conditional on extremely low sea ice summer seasons to establish a link between the sea ice lows and the thermodynamic and dynamic states of the atmosphere (Figure 6). During extremely low sea ice summers, positive temperature anomalies pre-exist over the Arctic during the beginning of the simulation period in February-March, especially from the Greenland to the Kara Seas and over the Canadian Archipelago (Figure 6(a)). Positive \acrshort{t2m} anomalies over the Arctic Ocean persist until spring before they largely disappear in June-July (Figure 6(a-d)). A reemergence of significant positive two metre temperatures occurs in August-September over the Arctic Ocean, though with a much smaller amplitude than at the beginning of the season (Figure 6(a,e)). Possible causes for the anomalously warm atmosphere in late winter and spring include enhanced poleward atmospheric heat and moisture transport \citep[e.g.][]{alekseev2019}, enhanced poleward oceanic heat transport \citep[e.g.][]{van2007bjerknes,jungclaus2010}, an anomalously strong conductive heat flux within an anomalously thin sea ice (see \cite{semtner1976} for a description of the thermodynamics of sea ice) and enhanced open water area related to negative sea ice area anomalies \citep{screen2010}. Since the positive \acrshort{t2m} anomalies in late winter and spring coincide with negative sea ice area and thickness anomalies (cf. Figures 5(a), 6(a-c) and Supplementary Figure S4), we expect that an increased upward surface sensible heat flux both due to enhanced upward conductive heat flux within an anomalously thin sea ice and due to reduced sea ice area contributes to the positive \acrshort{t2m} signal. The decline of the magnitude of the \acrshort{t2m} anomalies in summer is consistent with a constraint due to the climatological sea ice melting. Near-surface temperatures are at the freezing point and any additional heat input would accelerate sea ice melt, but not warm the air.  The weak positive \acrshort{t2m} signal in August-September could both be a consequence of an anomalously large amount of open water and a driver of sea ice reduction. 

Positive geopotential height anomalies pre-exist over the Arctic at the beginning of the simulation period, even though they are statistically significant only in a few grid points (Figure 6(f-g)). From April-May to May-June, they become instead statistically significant over large part of the Arctic (Figure 6(g-h)), while the positive \acrshort{t2m} anomaly signal at the surface is decreasing (Figure 6(b-c)). This suggests that anomalies in the atmospheric circulation in late spring and summer are not a pure consequence of the diabatic heating from the anomalously warm surface, but that on the contrary in this period it is the atmospheric dynamics that plays an active role in contributing to the reduction of sea ice.

\begin{figure*}[!htb]
    \vspace{0.6ex}
    \center{\includegraphics[width=\textwidth,trim={0.03cm 0.03cm 0.03cm 0.03cm},clip]
    {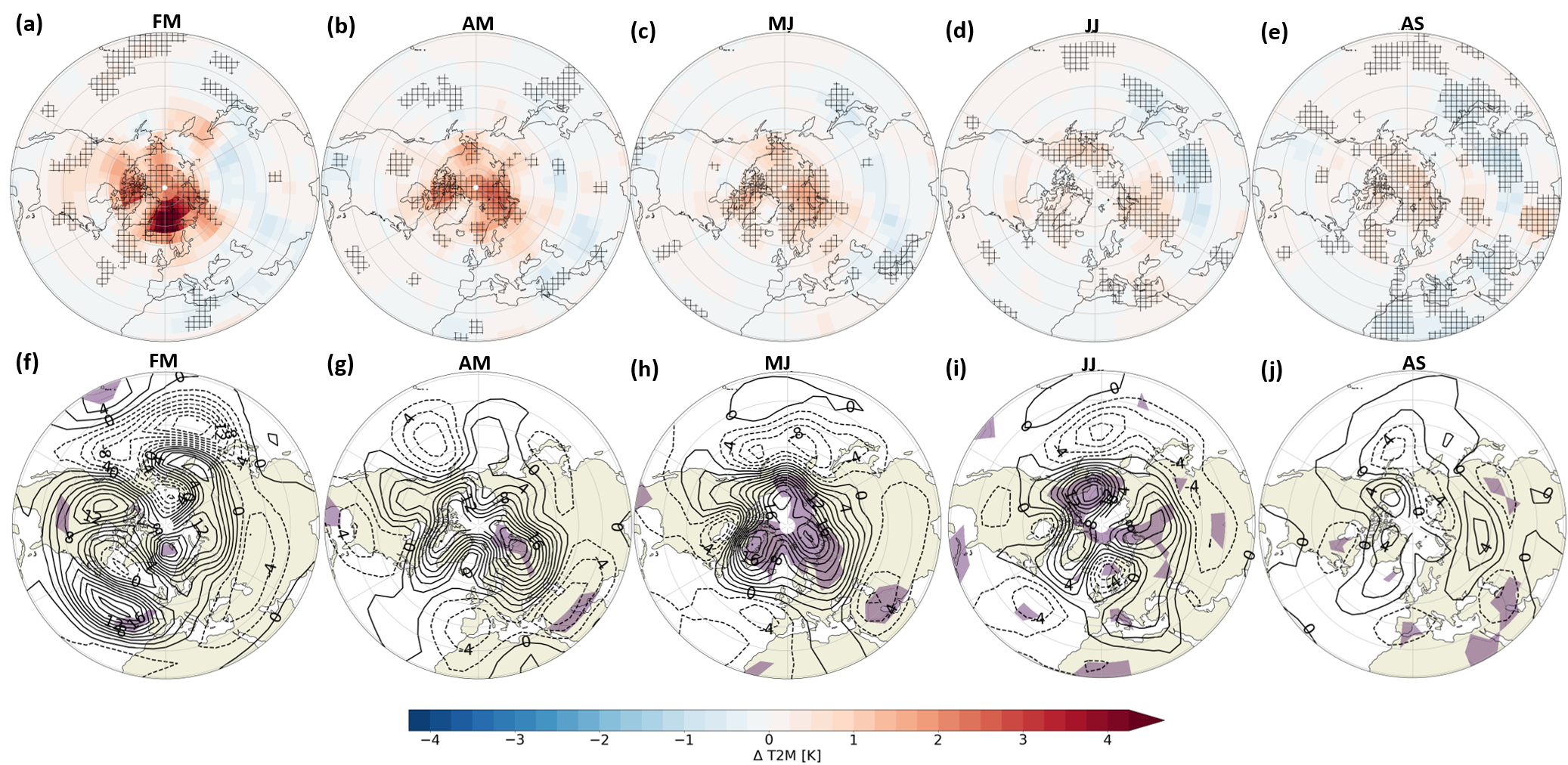}}
    \vspace*{-5mm}
    \caption{\label{} Composite mean (a-e) two metre temperature (T2M [K]) and (f-j) 500 hPa geopotential height (Z500 [gpm]; contour interval two gpm) anomalies conditional on extreme negative February-September mean pan-Arctic sea ice area anomalies presented as an average over the five rare event algorithm ensembles. The T2M and Z500 anomalies are averaged over (a,f) February-March (FM), (b,g) April-May (AM), (c,h) May-June (MJ), (d,i) June-July (JJ), (e,j) August-Septmeber (AS) and are respectively estimated relative to the climatology of the control ensembles. The sea ice area anomalies are classified as "extreme" if the sea ice area anomaly is smaller than -2.5 standard deviations of the control ensembles. The hatching in (a-e) and the shading in (f-j) denote statistical significance at the 5\% level according to a two-sided t-test.
    }
\end{figure*} 

\subsection{Surface energy budget and feedback mechanisms}\label{subsec3}
We perform a surface energy budget analysis for two purposes. Firstly, we are interested in the amount of anomalous energy accumulation in the snow-sea ice-ocean system that is related to surface-atmosphere energy flux anomalies between February and September. Secondly, we exploit the impact of the thermodynamic atmospheric state (e.g. temperature, water vapour, clouds) on the surface energy fluxes to understand how extreme negative pan-Arctic sea ice area anomalies are physically related to anomalous atmospheric conditions. We define the surface energy budget for an infinitesimally thin interface without heat storage located between the atmosphere and the snow-sea ice-ocean system (cf. \cite{serreze2014arctic}). The different terms of the budget are then given by     
\begin{equation}
    R_{SW}(1-\alpha) + R_{LW} + S + L  = -\epsilon \sigma T_s^4 + Q + M, 
\end{equation}

\noindent where $R_{SW}$ and $R_{LW}$ are the downward shortwave and downward longwave radiative fluxes, respectively; $\alpha$ is the surface albedo, $S$ and $L$ are the sensible and latent heat fluxes, $\epsilon$ is the surface emissivity, $\sigma=5.67 \cdot 10^{-8}\ $ W m\textsuperscript{-2} K\textsuperscript{-4} is the Stefan–Boltzmann constant and $T_s$ is the surface temperature. Q summarizes conductive and turbulent energy fluxes between the surface and the snow-sea ice-ocean system. M is the energy exchange associated with melting and freezing at the surface. Energy fluxes related to bottom sea ice growth, to temperature changes in the snow-sea ice-ocean system and the vertical energy flux from the ocean into the sea ice are implicitly included in Q. If not stated otherwise, we define upward (downward) energy fluxes and associated flux anomalies to be positive (negative). 

During summer seasons of extremely low pan-Arctic sea ice area, negative net surface-atmosphere energy flux anomalies occur over the Arctic Ocean and correspond to an anomalous energy accumulation within the snow-sea ice-ocean system (Figure 7(a)). The domain-average of these anomalies over all ocean grid boxes north of 70\degree N is -2.14 Wm\textsuperscript{-2}, which corresponds to a domain- and season- (February-September) integrated net energy accumulation of about 4.5 $\cdot$ 10\textsuperscript{20} J. This amount of energy would be sufficient to melt about 1.48 $\cdot$ 10\textsuperscript{3} km\textsuperscript{3} of sea ice, which is about three quarters of the February-September mean pan-Arctic sea ice volume anomaly being present during summer seasons of extremely low pan-Arctic sea ice area (Figure 5(b)). A direct contribution of the net accumulation of energy to the anomalies in the sea ice volume is, however, difficult for following reasons. One part of the energy is used to increase the temperature of the snow-sea ice-ocean system instead of being used for melting and an important part of the sea ice volume anomaly is already existing in late winter (Figure 5(b)). Likewise, the domain-average north of 70\degree N is limited to representing the vertical surface energy flux anomalies in the high Arctic, while the sea ice extends much further to the south during winter and spring. Nevertheless, the result indicates that the net February-September mean energy exchange between the atmosphere and the Arctic Ocean surface provides a substantial amount of energy potentially available to contribute to enhanced intra-seasonal sea ice reduction.

We are interested in the physical processes that drive the anomalous net energy transfer from the atmosphere to the Arctic Ocean during extremely low sea ice summers. For this purpose, we investigate how the net atmosphere-surface energy flux anomalies are partitioned into their different components over the course of the season (Figure 7(b-c)). Both the net atmosphere-surface energy flux anomalies and their different components exhibit a pronounced seasonality. An enhanced net energy transfer from the atmosphere to the snow-sea ice-ocean system occurs between April-May and August-September, which includes the period of anomalous intra-seasonal reduction in the pan-Arctic sea ice area and sea ice volume (cf. Figures 7(b) and 5). In contrast, slightly positive net atmosphere-surface energy flux anomalies indicate a small anomalous release of energy from the Arctic Ocean to the atmosphere in late winter (Figure 7(b)). The latter is consistent with a contribution of preconditioning via the winter sea ice-ocean state to extremely low \acrshort{febsep} pan-Arctic sea ice area (see section 3.2.2). 

Regarding the entire season, the net atmosphere-surface energy flux anomalies are dominated by the radiative (net shortwave and net longwave radiation) compared to the non-radiative (sensible and latent heat) fluxes (Figure 7(c)). Slightly positive non-radiative flux anomalies explain the small anomalous energy transfer from the Arctic Ocean to the atmosphere in late winter, while negative sensible heat flux anomalies in May-June suggest a direct contribution of anomalously high air temperatures to enhanced snow-sea ice-ocean warming or to enhanced snow-sea ice melt. The negative sensible heat flux anomalies in May-June coincide with the appearance of a statistically significant positive \acrshort{z500} anomaly pattern and thus with an anomalously warm state of the lower troposphere (cf. Figures 7(c) and 6). A strong contribution of net shortwave radiative fluxes to the seasonal accumulation of energy in the Arctic Ocean occurs between April-May and August-September. The net longwave radiative flux anomalies are smaller in magnitude than the shortwave ones, but still contribute to a statistically significant anomalous accumulation of energy in the Arctic Ocean between March-April and June-July. An enhanced net longwave radiative forcing on the Arctic sea ice during spring is consistent with \cite{kapsch2013,kapsch2019}. \cite{kapsch2013,kapsch2019} establishes a link between observed extremely low September Arctic pan-Arctic sea ice area and an anomalous early melt onset due to the footprint of an anomalous cloudy and moist atmosphere on the net longwave radiative fluxes.

\begin{figure*}[!htb]
    \vspace{0.6ex}
    \center{\includegraphics[width=\textwidth,trim={0.03cm 0.03cm 0.03cm 0.03cm},clip]
    {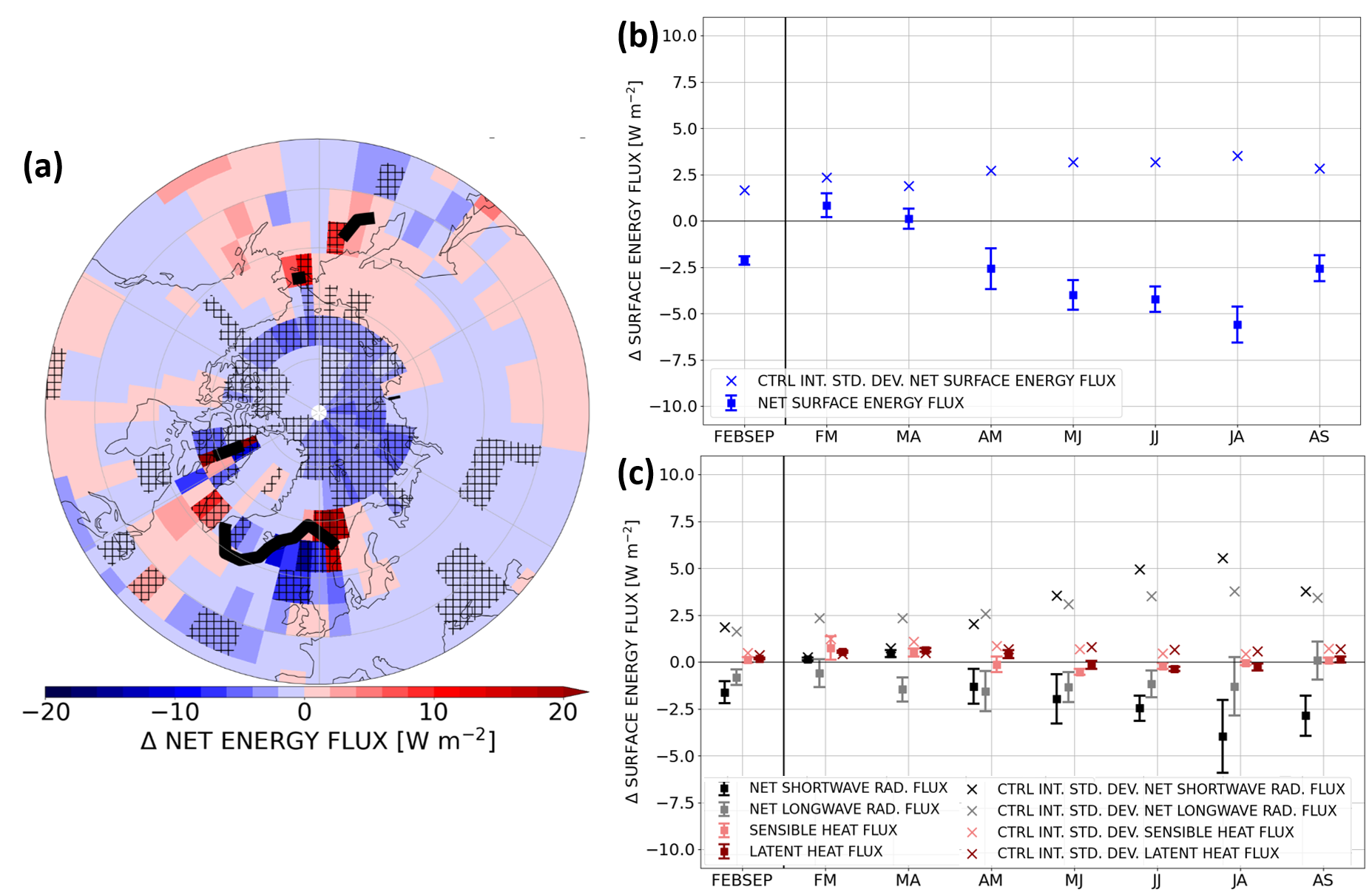}}
    \vspace*{-5mm}
    \caption{\label{} Rare event algorithm experiments with a resampling time of $\tau_r = 5$ days: Composite mean surface energy flux [Wm\textsuperscript{-2}] anomalies conditional on extreme negative February-September mean pan-Arctic sea ice area anomalies equal or smaller than -2.5 standard deviations of the control ensembles. Mean anomalies are presented as an average over the five rare event algorithm experiments and are evaluated relative to the control run. Hatching in (a) indicates statistical significance at the 5\% level and error bars in (b-c) show 95\% confidence intervals obtained from the statistics of the five estimates (see Supplementary Information S1). (a) February-September mean net surface-atmosphere energy flux anomalies (sensible + latent + net longwave + net shortwave). (b-c) February-September and bimonthly mean domain-averaged (b) net surface-atmosphere energy flux anomalies and (c) surface sensible and latent heat flux and net shortwave and net longwave radiative flux anomalies. The averaging is performed over all ocean grid boxes north of 70\degree N. The "x" markers respectively indicate the climatological standard deviations in the control run. The black contour line in (a) indicates the climatological February-September mean sea ice edge defined as the 15\% sea ice concentration line.}
\end{figure*} 

In order to trace back the net radiative flux anomalies to anomalous atmospheric conditions, we subdivide the net shortwave and longwave radiative surface fluxes into their downward and upward components (Figure 8(a,b)). Strongly enhanced downward longwave radiation occurs between late winter and late spring, which is partly balanced by enhanced upward longwave radiation in accordance with anomalously high (near-)surface air temperatures during that period (cf. Figures 8(a) and 6(a-e)). In June-July and July-August, positive downward radiative flux anomalies are much smaller than in spring, but counteracting upward longwave radiative flux anomalies drop to zero because sea ice-snow melting constraints surface temperatures to be close to the freezing point. In August-September, weakly enhanced upward and downward longwave radiative flux anomalies compensate each other. Regarding the shortwave radiative fluxes, reduced downward fluxes are overcompensated by reduced upward fluxes, leading to an anomalous net accumulation of energy in the snow-sea ice-ocean system due to shortwave fluxes between April-May and August-September (Figure 8(b)).

The combination of enhanced downward longwave and reduced downward shortwave radiation suggests that extremely low sea ice summers are characterized by enhanced cloudiness. During extremely low sea ice summers, statistically significant positive cloud cover anomalies are present from late winter to late spring (Figure 8(c)). The enhanced cloud cover is accompanied by positive integrated water vapour anomalies being statistically significant from late winter to mid-summer (Figure 8(c)). Consequently, apart from anomalously high air temperatures, enhanced cloudiness and water vapour in the atmosphere contribute to the positive downward and net longwave radiative surface fluxes during extremely low sea ice summers. Reduced upward shortwave radiative fluxes are naturally emerging due the reduction in the downward shortwave radiative fluxes. However, a strongly reduced surface albedo due to reduced sea ice cover, i.e. the sea ice-albedo feedback, likewise contributes to the accumulation of energy related to the reduced upward shortwave fluxes (Figure 8(d)).    
\begin{figure*}[!htb]
    \vspace{0.6ex}
    \center{\includegraphics[width=\textwidth,trim={0.03cm 0.03cm 0.03cm 0.03cm},clip]
    {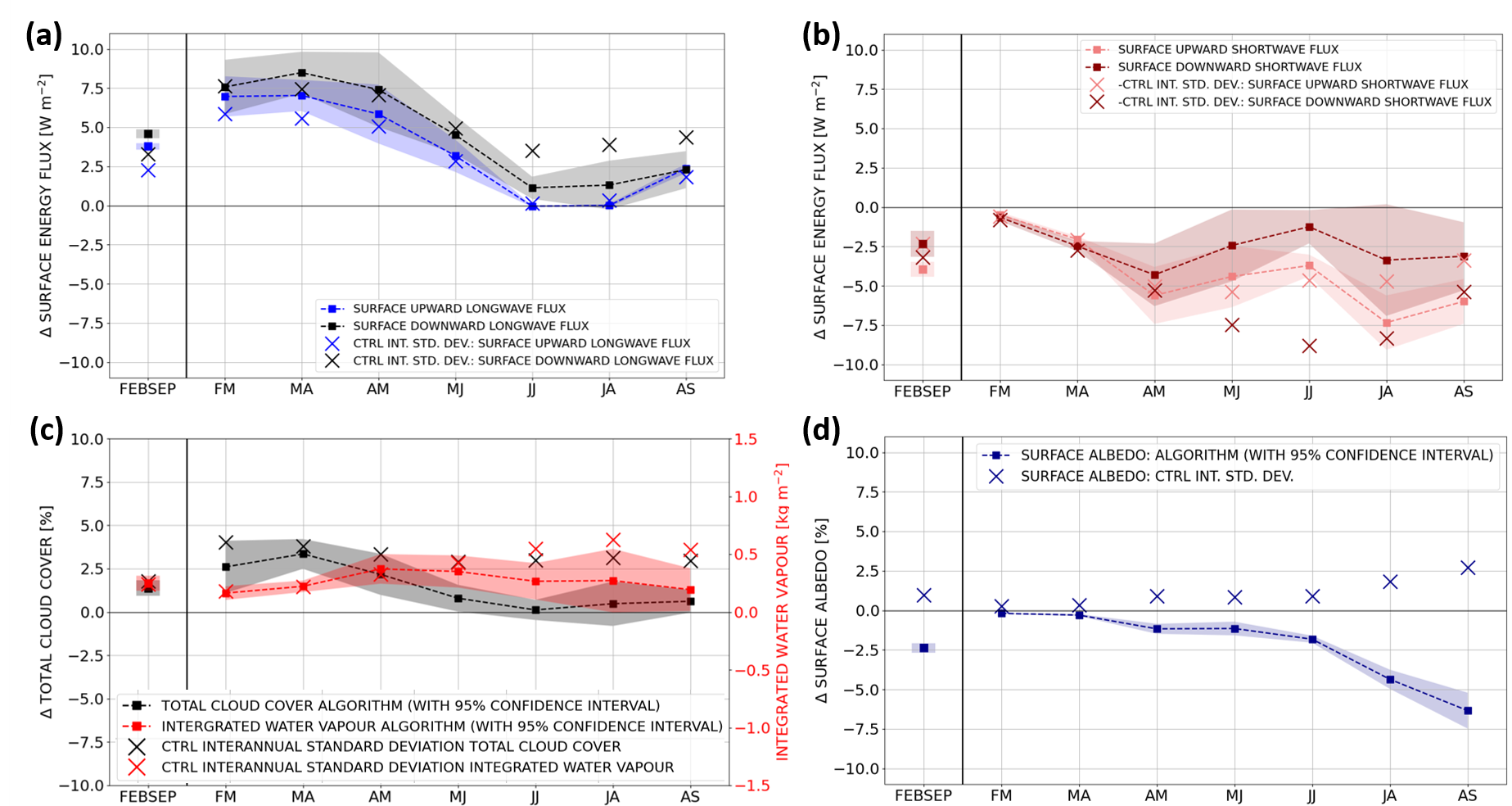}}
    \vspace*{-5mm}
    \caption{\label{} Rare event algorithm experiments with a resampling time of $\tau_r = 5$ days: February-September and bimonthly mean domain-averaged anomalies of different variables conditional on extreme negative February-September mean pan-Arctic sea ice area anomalies equal or smaller than -2.5 standard deviations of the control ensembles. Mean anomalies are presented as an average over the five rare event algorithm experiments and are evaluated relative to the control run. Shading indicates 95\% confidence intervals obtained from the statistics of the five estimates (see Supplementary Information S1). The domain is defined as in Figure 7. (a-b) Surface upward and downward (a) longwave and (b) shortwave radiative flux anomalies. Direction-independent absolute values of the downward and upward fluxes are considered, i.e., a positive (negative) anomaly indicates a radiative flux that is stronger (weaker) in magnitude than the climatology. (c) (black) Total cloud cover [\%] and (red) integrated water vapour [kg m\textsuperscript{-2}] and (d) surface albedo [\%] anomalies. (a-d) The "x" markers indicate respectively the climatological standard deviations in the control run.}
\end{figure*} 

\section{Discussion and conclusions}\label{sec5}

The present study demonstrates the applicability of a rare event algorithm to improve the sampling efficiency of extremely low pan-Arctic sea ice area during the melting season. The simulations with the rare event algorithm produce several hundreds of times more extremes than the control run for the same computational cost, and allow us to compute return times of extremes two to three orders of magnitude larger than feasible with direct sampling. Owing to the rare event algorithm, we compute statistically significant composites of dynamical quantities conditional on the occurrence of extremely low sea ice summers with return times of more than 200 years. In this way, we identify four processes and conditions that lead to extreme negative February-September mean pan-Arctic sea ice area anomalies in \acrshort{plasim-lsg}: 1) preconditioning through the winter sea ice-ocean state, 2) enhanced downward longwave radiation due to an anomalously moist and warm Arctic atmosphere in spring, 3) enhanced downward sensible heat fluxes in May-June and 4) the sea ice-albedo feedback becoming active throughout late spring and summer.

The preconditioning manifests in large negative winter pan-Arctic sea ice volume anomalies as a result of both negative sea ice area and sea ice thickness anomalies (Figure 5 and Supplementary Figure S4(a,b)). The preconditioning of anomalously low summer pan-Arctic sea ice area through negative winter sea ice thickness anomalies is in agreement with \cite{chevallier2012role}, \cite{holland2011inherent} and \cite{kauker2009adjoint}, who emphasize the important role of winter-spring sea ice thickness for the probability of sea ice to survive the melt season. \cite{chevallier2012role} further argues that the winter-spring sea ice thickness, in particular the area covered by sea ice thicker than a critical thickness, is a better predictor of the summer pan-Arctic sea ice area than the winter-spring sea ice area itself. Anomalous states in the winter-spring sea ice thickness and sea ice area can be generated by a variety of mechanisms, including the dynamical forcing of the atmospheric circulation on the sea ice \cite[e.g.][]{ogi2016,rigor2002response}, mechanisms related to the sea ice memory itself \citep{blanchard2011persistence} and an anomalous heat content in the ocean \citep{polyakov2012,comiso2012}. While a future study is required to examine the precise cause of the preconditioning in \acrshort{plasim-lsg}, direct dynamical causes can be excluded due to the fact that a purely thermodynamic sea ice model is used in the present work.
  
The second process is related to an anomalously moist and warm atmosphere, which manifests in enhanced downward longwave radiation between February-March and July-August accompanied by enhanced downward sensible heat flux in May-June. Enhanced spring downward longwave radiation as one driver of extremely low summer Arctic sea ice area is consistent with \cite{kapsch2013, kapsch2019}, who demonstrate a linkage between observed extremely low September pan-Arctic sea ice area and a persistent anomalously moist and cloudy atmosphere in the preceding spring. \cite{kapsch2019} and \cite{persson2012} argue that enhanced downward longwave radiation in spring can lead to enhanced surface warming prior to the melt onset and to an earlier melt onset compared to climatology. \cite{kapsch2013, kapsch2019} further demonstrate that the observed anomalously moist spring atmosphere prior to extremely low September Arctic sea ice area is related to enhanced meridional water vapour transport. In \acrshort{plasim-lsg}, the anomalously moist and warm atmosphere may both be a trigger, i.e. via enhanced meridional heat and water vapour transport into the Arctic, and a response of extreme negative sea ice area anomalies, i.e. due to enhanced evaporation and sensible heat loss from the Arctic Ocean to the atmosphere due to positive surface temperature and open water fraction anomalies.

The third process suggests that the atmosphere may act not only as an amplifier but also as a trigger of sea ice retreat. In May-June, enhanced downward sensible heat fluxes over the Arctic Ocean are associated with a reinforcement of positive 500 hPa geopotential height anomalies compared to earlier months, while positive \acrshort{t2m} anomalies are decreasing in amplitude (cf. Figures 6 and 7(c)). Consequently, the positive 500 hPa geopotential height anomalies and thus the anomalously warm lower troposphere during the spring-summer transition cannot be explained alone by a diabatic forcing through an anomalously warm surface. The atmospheric dynamics and energy fluxes in May-June are compatible with the characteristics of midlatitude heatwaves. The possible role of Arctic heatwaves as drivers of extreme sea ice retreat is relatively understudied in the literature, and a detailed analysis of their impacts and dynamical origin will be subject of a future study.

The fourth process is given by the sea ice-albedo feedback, which sets in during April-May as a consequence and amplifier of extremely low sea ice cover. Compared to the different atmosphere-surface flux components discussed in section 3.3, the sea ice-albedo feedback explains the largest amount of anomalous Arctic Ocean net energy accumulation during extremely low sea ice summers. Future rare event algorithm simulations with a prescribed surface albedo are a possibility to quantify the relative contribution of the sea ice-albedo feedback to extreme negative pan-Arctic sea ice area anomalies.   

Overall, two reasons let us conclude that, in \acrshort{plasim-lsg}, the preconditioning through the winter-spring sea ice-ocean state is a more dominant driver of extreme negative February-September mean pan-Arctic sea ice area anomalies than the anomalous atmospheric conditions. Firstly, pan-Arctic sea ice volume anomalies prior to extremely low sea ice summers are about one and a half standard deviations away from the climatology (Figure 5(b)), while anomalies of atmospheric quantities and of downward atmosphere-surface energy fluxes are mostly less than one standard deviation away from the climatology (Figures 7(c) and 8(a-c)). Secondly, the most extreme February-September mean sea ice area anomalies obtained with the 5-days resampling time experiments only have marginally larger amplitudes than the ones obtained with the 30-days resampling time experiments (Figure 3). Consequently, the targeted sampling of atmospheric drivers of sea ice reduction that act on the upper range of synoptic and on submonthly time scales do not substantially increase the magnitude of the most extreme sea ice area anomalies compared to the 30-days resampling time experiments, which are primarily designed to efficiently sample oceanic drivers of low sea ice states. We highlight, however, that this study is based on a relatively low resolution climate model with a purely thermodynamic sea ice model. A direct dynamic forcing of the atmospheric circulation on the sea ice, e.g. related to synoptic-scale storms or to the \acrshort{ao} and the \acrshort{ad} pattern, are therefore not captured by the model.

We point out that, even though the experiments with the rare event algorithm presented in this paper are designed to improve the sampling efficiency of February-September mean pan-Arctic sea ice area anomalies, the method indirectly also oversamples trajectories characterized by negative anomalies of the pan-Arctic sea ice area during the annual sea ice minimum in September (Figure 2(a,b); note that the correlation coefficient between \acrshort{febsep} and \newacronym{sep}{SEP}{September} \acrfull{sep} pan-Arctic sea ice area in the control run is 0.69). This suggests a potential applicability of the rare event algorithm to study extremes of the annual Arctic sea ice minimum. This could also be obtained by changing the observable used to weight the trajectories from the pan-Arctic sea ice area to its time derivative. Finally, we highlight that in the experiments presented in this paper the initial conditions of the ensemble have been taken from the control run in order to have an unbiased sampling of the invariant measure of the dynamics. In this way, the return times and statistics computed with the algorithm are related to unconditional probability distributions. A different type of experiment would consist in starting each ensemble member from the same initial condition, as done for ensemble weather and seasonal climate predictions. In this case, the algorithm would give access to the statistics of extreme events conditional on the chosen initial condition. This approach could be of great interest in the context of seasonal predictions in order to estimate the risk of observing a record low of Arctic sea ice in summer given the conditions at the beginning of the melting season. This strategy and the extent to which the accuracy of statistical estimators based on the importance sampling formula in these type of experiments improves the results compared to direct sampling constitute an exciting line of research that will be explored in future works.

\mbox{~}
\newpage
\mbox{~}
\newpage
\bibliography{References_JSauer}

\section*{Statements \& Declarations}

\noindent\textbf{Competing Interests} \\
The authors have no relevant financial or non-financial interests to disclose. \\

\noindent\textbf{Author Contributions} \\
All authors contributed to the study conception and design. Model simulations, data collection and analyses were performed by the first author Jerome Sauer. The first draft of the manuscript was written by the first author Jerome Sauer and all authors commented on previous versions of the manuscript. All authors read and approved the final manuscript. \\  

\noindent\textbf{Data Availability} \\
The data required to reproduce the results of this paper are freely available on the Zenodo platform \citep{sauer2023}. \\

\noindent\textbf{Funding} \\
This publication is supported by the FSR Seedfund program and by the French Community of Belgium as part of a FRIA (Fund for research training in industry and agriculture) grant.

\section*{Acknowledgements}
Computational resources have been provided by the supercomputing facilities of the Université catholique de Louvain (CISM/UCL) and the Consortium des Équipements de Calcul Intensif en Fédération Wallonie Bruxelles (CÉCI) funded by the Fond de la Recherche Scientifique de Belgique (F.R.S.-FNRS) under convention 2.5020.11 and by the Walloon Region. This project is supported by the Belgian Science Policy (BELSPO) project RESIST (CONTRAT N° RT/23/RESIST). This project is also supported by the European Union (ERC, ArcticWATCH, 101040858). Views and opinions expressed are however those of the author(s) only and do not necessarily reflect those of the European Union or the European Research Council Executive Agency. Neither the European Union nor the granting authority can be held responsible for them. François Massonnet is a F.R.S.-FNRS Research Associate. GZ has been supported by the “Programma di Ricerca in Artico” (PRA; project no. PRA2019-0011, SENTINEL).

\mbox{~}
\newpage
\section*{Supplementary Information}
\setcounter{figure}{0}
\renewcommand{\thefigure}{S\arabic{figure}} 

\subsection*{S1: Statistical significance}
We use a two-sided t-test to estimate the statistical significance of composite maps and we use the t-distribution of construct confidence intervals. We use the 5\% significance level. We subdivide the control run in several ensembles and perform multiple experiments with the rare event algorithm. We denote $M$ the number of independent experiments in which extreme events of interest occur. We compute the t-value as $t=\sqrt{M} \frac{\mathbb{E}_0(f)}{\sigma(f)}$, where $\mathbb{E}_0(f)$ is the empirical average of a quantity, e.g. the average over composite mean values obtained from multiple experiments, and $\sigma(f)$ is the empirical standard deviation of that quantity, e.g. the standard deviation over composite mean values obtained from multiple experiments. We denote a composite mean value $\mathbb{E}_0(f)$ as statistically significant at the 5\% level if the absolute value of t is larger than 2.776 for $M=5$, 3.182 for $M=4$ and 4.303 for $M=3$. We likewise construct 95\% confidence intervals from the student's t-distribution. In the present work, we have $M=5$ for all composite analyses and $M \in \{3, 4, 5\}$ for the return time estimates. The variable sample size regarding the return times estimated with the rare event algorithm is due to the fact that the rare event algorithm experiments with different biasing parameters $k$ cover slightly different ranges of pan-Arctic sea ice area anomalies. Hence, the estimates of the return times and of their confidence intervals is based on the output of five rare event algorithm experiments in a broad range of sea ice area anomalies and on the output of three or four experiments in the upper and lower amplitude range of sea ice area anomalies.

\subsection*{S2: Adequacy of the rare event simulations}
One problem potentially occurring in rare event simulations is degeneracy. Degeneracy means that the entire ensemble is effectively shrinked to a small number of ensemble members that are distinguishable only in a small last part of the simulation. The degeneracy level in the simulations performed for this work, however, is relatively low (Figures S5 and S6). For each resampling step, the fraction of surviving trajectories with respect to the ensemble size is larger than 90\% for the 5-days resampling time and between about 60\% and 90\% for the 30-days resampling time experiments (Figure S5(a-b); note that the decrease of the survival fraction in July consistently coincides with an increase in the interannual variability of sea ice area anomalies (cf. Figure S5(a-b) and Figures 1(c), 2(a,b) of the main text)). The comparatively high rate of surviving trajectories per resampling step results in a relatively large number of different initial-time ancestors that lead to at least one final-time decendant (between 51 and 133; Figure S5(c-d)). Furthermore, the number of different initial-time ancestors that lead to at least one (or more) extreme negative February-September pan-Arctic sea ice area anomaly equal or smaller than -2.5 standard deviations of the control ensembles is substantially larger than the number of extreme events sampled in the control ensembles (Figure S5(c-d); note that the total number of extreme events counted over partially or fully overlapping trajectories in the rare event simulations is in the order of several hundreds per ensemble). Finally, the low degeneracy level in the rare event simulations becomes evident from the distribution of the number of final-time decendants that are associated with the different initial-time ancestors. Most of the initial-time ancestors lead to between 1 and less than 50 final-time decendants (Figure S6). The present study does not contain a situation where one single initial-time ancestor dominates the entire ensemble.  

\begin{figure*}[!htb]
    \vspace{0.6ex}
    \center{\includegraphics[width=\textwidth,angle=0,trim={0.03cm 0.03cm 0.03cm 0.03cm},clip]
    {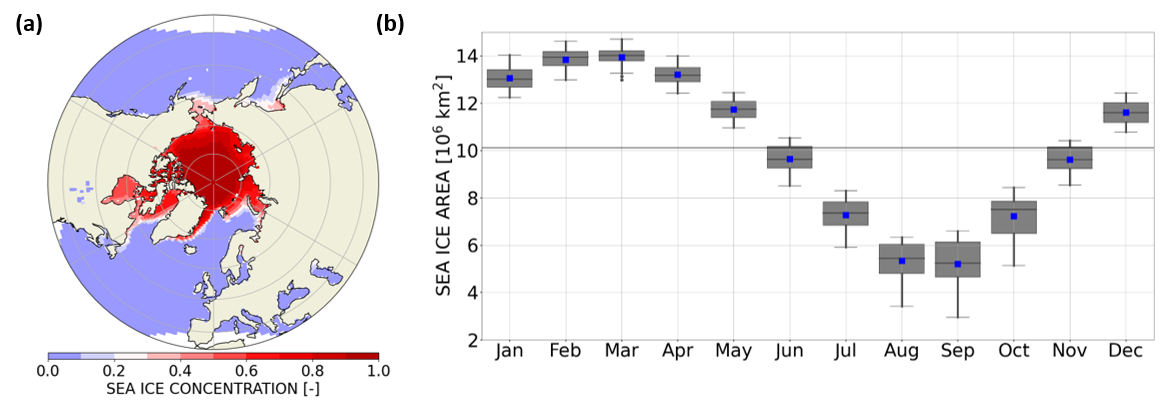}}
    \vspace*{-5mm}
    \caption{\label{} (a) 1979-2015 mean \newacronym{sic}{SIC}{sea ice concentration}\acrfull{sic} from \newacronym{eumetsat}{EUMETSAT}{European Organisation for the Exploitation of Meteorological Satellites} \acrfull{eumetsat} \newacronym{osisaf}{OSI SAF}{Ocean and Sea Ice Satellite Application Facility} \acrfull{osisaf} data \citep{eumetsat1} and (b) monthly mean pan-Arctic sea ice area [10\textsuperscript{6} km\textsuperscript{2}] from \acrshort{eumetsat} \acrshort{osisaf} data between 1979 and 2015 \citep{eumetsat2}. The horizontal gray line shows the annual mean pan-Arctic sea ice area.}
\end{figure*} 

\begin{figure*}[!htp]
    \vspace{0.6ex}
    \center{\includegraphics[width=\textwidth,trim={0.03cm 0.03cm 0.03cm 0.03cm},clip]
    {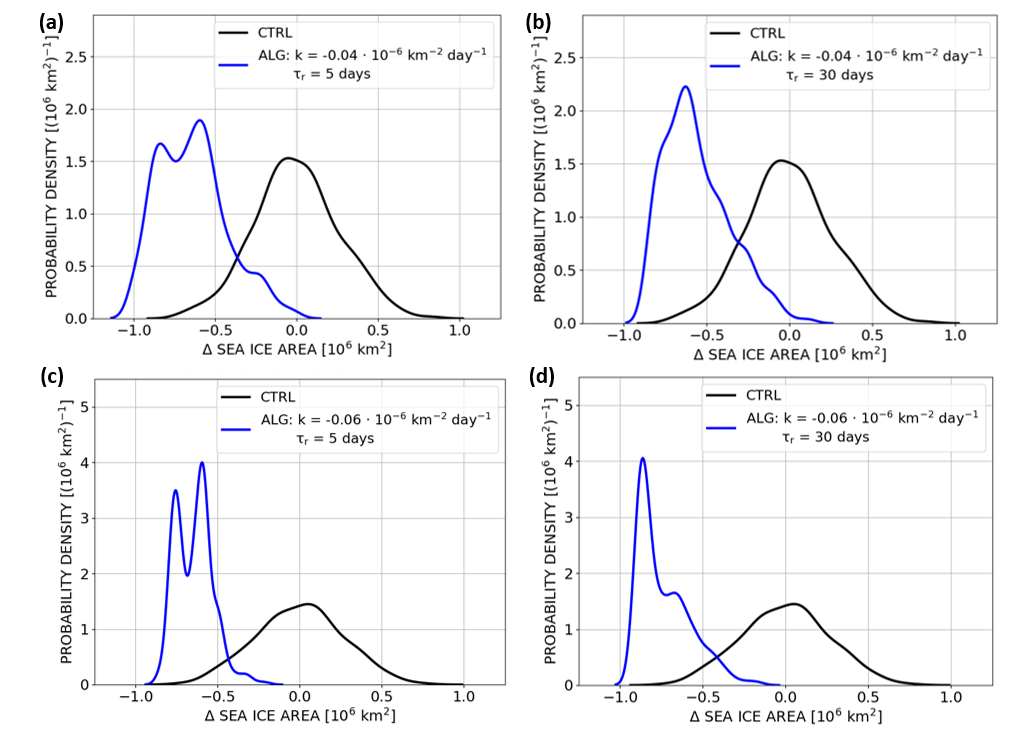}}
    \vspace*{-5mm}
    \caption{\label{} Probability distribution functions of February-September mean pan-Arctic sea ice area anomalies relative to the control climatology for (black) the control run itself and (blue) experiments with the rare event algorithm for (a,c) a resampling time of 5 days and (b,d) a resampling time of 30 days. (a,b) Control ensembles three and five and rare event algorithm experiments corresponding to ensembles three and five, i.e. with $k=-0.04\ \cdot\ $ 10\textsuperscript{-6} km\textsuperscript{-2} day\textsuperscript{-1}. (c,d)  Control ensemble one and rare event algorithm experiment corresponding to ensemble one, i.e. with $k=-0.06\ \cdot$ 10\textsuperscript{-6} km\textsuperscript{-2} day\textsuperscript{-1}.}
\end{figure*} 

\begin{figure*}[!htp]
    \vspace{0.6ex}
    \center{\includegraphics[width=\textwidth,trim={0.03cm 0.03cm 0.03cm 0.03cm},clip]
    {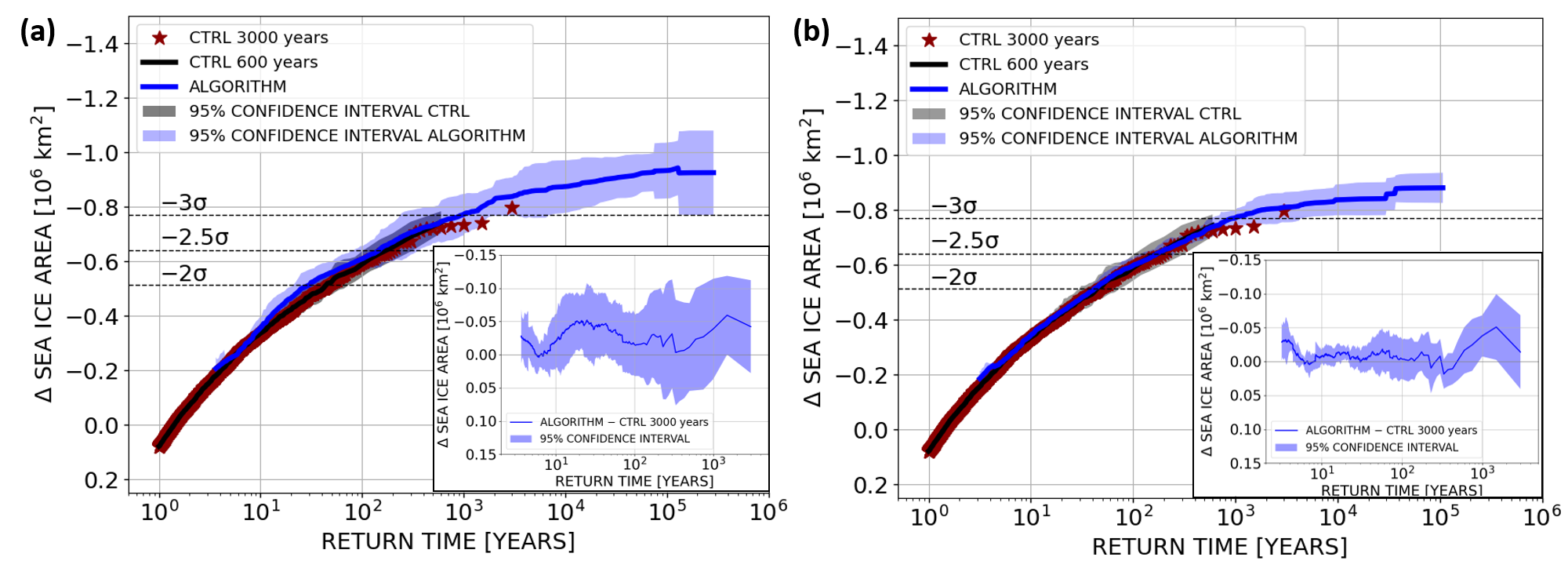}}
    \vspace*{-5mm}
    \caption{\label{} Return curves for February-September mean pan-Arctic sea ice area anomalies relative to the control run. (Red stars) The direct estimate of return times from the 3000 year control run, (black line) the average estimate over the five 600-member ensembles of the control run and (blue line) the average estimate over the overlap of at least three out of five 600-member rare event algorithm experiments. Shading denotes the 95\% confidence interval obtained from the statistics of the three to five estimates assuming a student's t-distributed estimator (see Supplementary Information S1). (a) Rare event algorithm with 5-days resampling time and (b) with 30-days resampling time. The dashed lines in (a) and (b) indicate anomalies of minus two, minus two and a half and minus three standard deviations with respect to the control run. (a-b) The insets show the difference between the return level estimates obtained with the rare event algorithm and the ones directly estimated from the 3000 year control run together with the 95\% confidence intervals of these differences (the confidence intervals of the differences neglect the uncertainty of the control estimates).}
\end{figure*}

\begin{figure*}[!htb]
    \vspace{0.6ex}
    \center{\includegraphics[width=\textwidth,trim={0.03cm 0.03cm 0.03cm 0.03cm},clip]
    {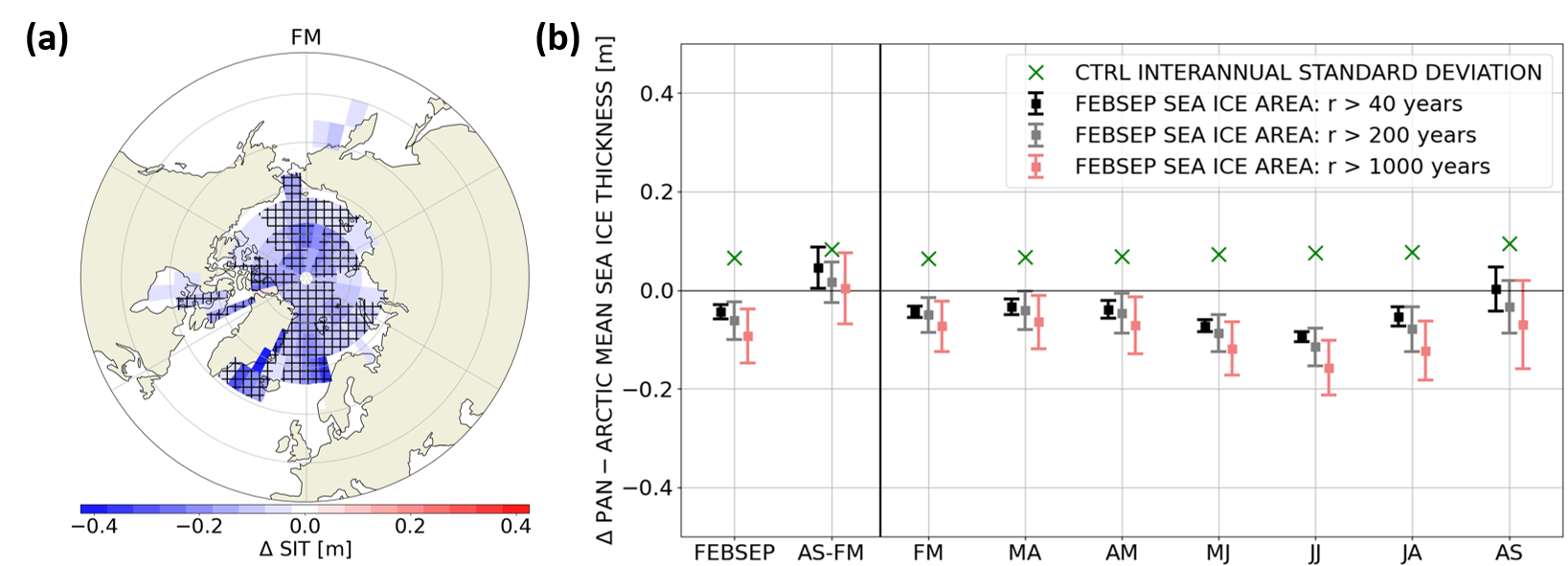}}
    \vspace*{-5mm}
    \caption{\label{} Composite mean \newacronym{sit}{SIT}{sea ice thickness}\acrfull{sit} anomalies [m] conditional on extreme negative February-September mean pan-Arctic sea ice area anomalies equal or smaller than (a) -2.5 standard deviations and (b) (black) -2 standard deviations, (gray) -2.5 standard deviations and (red) -3 standard deviations of the control run (roughly corresponds to extremes with return times of more than 40, 200 and 1000 years). (a,b) The composite estimates are presented as an average over the five rare event algorithm ensembles and the hatching in (a) and error bars in (b) represent statistical signficance at the 5\% level and 95\% confidence intervals respectively (see Supplementary Information S1). The anomalies are computed with respect to the control climatologies of the five control ensembles. (a) Spatial field of \newacronym{fm}{FM}{February-March}\acrfull{fm} \acrshort{sit} anomalies and (b) spatial average of mean \acrshort{sit} anomalies over all grid box north of 40\degree N conditional on the presence of sea ice. "AS-FM" denotes the difference between \newacronym{as}{AS}{August-September} \acrfull{as} and \acrshort{fm} and the green "x" markers indicate the climatological standard deviation in the control run.}
\end{figure*} 

\begin{figure*}[!htb]
    \vspace{0.6ex}
    \center{\includegraphics[width=\textwidth,trim={0.03cm 0.03cm 0.03cm 0.03cm},clip]
    {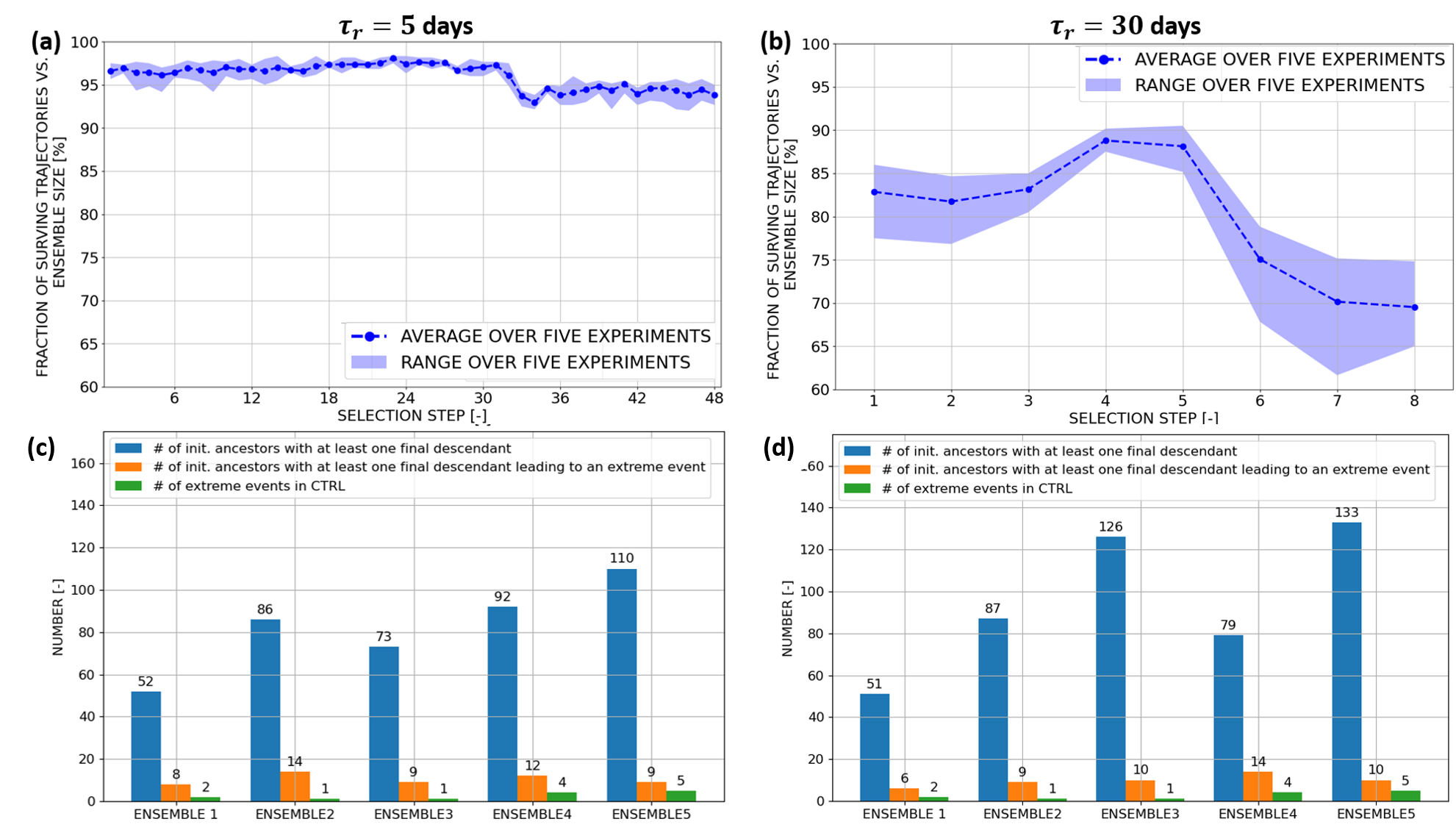}}
    \vspace*{-5mm}
    \caption{\label{} (a-b) Fraction of surviving trajectories with respect to the ensemble size for each selection step [\%] shown as (markers) the average and (shading) the minimum-maximum range over five experiments with the rare event algorithm. (c-d) Number of (blue) different initial-time ancestors with a non-zero number of final-time decendants, (orange) different initial-time ancestors with a non-zero number of final-time decendants leading to an extreme negative February-September mean pan-Arctic sea ice area anomaly equal or smaller than -2.5 standard deviations of the control ensembles and (green) the number of extreme events in the control ensembles (same extreme event definition as for (orange)). (a,c) show the experiments with a resampling time of 5 days and (b,d) the ones with a resampling time of 30 days.}
\end{figure*} 

\begin{figure*}[!htb]
    \vspace{0.6ex}
    \center{\includegraphics[width=\textwidth,trim={0.03cm 0.03cm 0.03cm 0.03cm},clip]
    {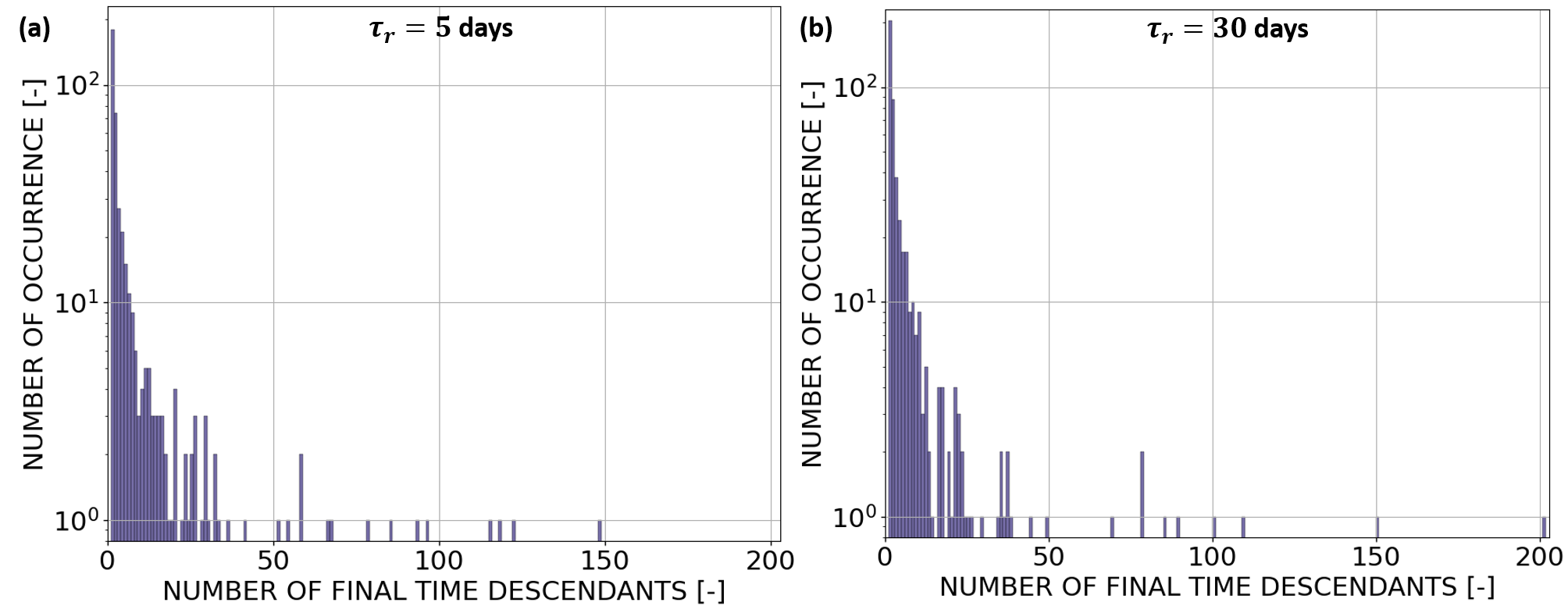}}
    \vspace*{-5mm}
    \caption{\label{}: Histograms of the number of final-time decendants for the different initial-time ancestors presented in Figure S5(c-d) aggregated over the five experiments with a resampling time of (a) 5 days and (b) 30 days.}
\end{figure*} 

\end{document}